\documentclass[dvipdfm]{aa}
\usepackage[dvips]{graphicx}
\usepackage{txfonts}
\usepackage{natbib}
\bibpunct{(}{)}{;}{a}{}{,}

\begin{document}

\title{A Sino-German $\lambda$6~cm polarization survey of the Galactic plane} 
\subtitle{III. The region from $10\degr$ to $60\degr$ longitude}

\author{X. H. Sun\inst{1,2}
        \and W. Reich\inst{2} 
        \and J. L. Han\inst{1}
        \and P. Reich\inst{2}
        \and R. Wielebinski\inst{2}
        \and C. Wang\inst{1}
        \and P. M\"uller\inst{2}}

\institute{National Astronomical Observatories, CAS, Jia-20 Datun Road, 
           Chaoyang District, Beijing 100012, China\\
           \email{[xhsun;hjl;wangchen]@nao.cas.cn}
            \and Max-Planck-Institut f\"{u}r Radioastronomie, 
                 Auf dem H\"ugel 69, 53121 Bonn, Germany\\
           \email{[xhsun;wreich;preich;rwielebinski;pmueller]@mpifr-bonn.mpg.de}}

\date{Received / Accepted}

\abstract
{} 
{We aim to study polarized short-wavelength emission from the inner Galaxy, 
which is nearly invisible at long wavelengths because of depolarization. 
Information on the diffuse continuum emission at short wavelengths is required 
to separate Galactic thermal and non-thermal components including existing 
long-wavelength data.
}
{We have conducted a total intensity and polarization survey of the Galactic 
plane at $\lambda$6~cm using the Urumqi 25 m telescope for the Galactic 
longitude range of $10\degr\leq l\leq60\degr$ and the Galactic latitude range 
of $|b|\leq5\degr$. Missing absolute zero levels of Stokes $U$ and $Q$ maps 
were restored by extrapolating the WMAP five-year K-band polarization data. 
For total intensities we recovered missing large-scale components by referring 
to the Effelsberg $\lambda$11~cm survey. 
}
{Total intensity and polarization maps are presented with an angular resolution
of $9\farcm5$ and a sensitivity of 1~mK~$T_{\rm B}$ and 0.5~mK~$T_{\rm B}$ in
total and polarized intensity, respectively. The $\lambda$6~cm polarized 
emission in the Galactic plane originates within about 4~kpc distance, which 
increases for polarized emission out of the plane. The polarization map shows 
``patches", ``canals" and ``voids" with no correspondence in total intensity. 
We attribute the patches to turbulent magnetic field cells. Canals are caused 
by abrupt variation of polarization angles at the boundaries of patches rather 
than by foreground Faraday Screens. The superposition of foreground and Faraday 
Screen rotated background emission almost cancels polarized emission locally, 
so that polarization voids appear. By modelling the voids, we estimate the 
Faraday Screen's regular magnetic field along the line-of-sight to be larger 
than about 8~$\mu$G. We separated thermal (free-free) and non-thermal 
(synchrotron) emission according to their different spectral indices. The 
spectral index for the synchrotron emission was based on WMAP polarization data.
The fraction of thermal emission at $\lambda$6~cm is about 60\% in the plane. 
} 
{The Sino-German $\lambda$6~cm polarization survey of the inner Galaxy provides 
new insights into the properties of the magnetized interstellar medium for this 
very complex Galactic region, which is Faraday thin up to about 4~kpc in the 
Galactic plane. Within this distance polarized patches were identified as 
intrinsic structures related to turbulent Galactic magnetic fields for spatial 
scales from 20(d/4~kpc) to 200(d/4~kpc)~pc
}


\maketitle

\section{Introduction}

We have mapped a broad region of the Galactic plane at $\lambda$6~cm as part of 
the Sino-German polarization survey of the Milky Way. In previous papers 
(\citealt{shr+07}, Paper~I; \citealt{grh+10}, Paper~II) we studied the Galactic 
plane emission in the anti-centre direction of the Milky Way. They included the 
study of the properties of selected regions showing evidence for strong Faraday 
rotation. Even in the direction of the anti-centre Faraday Screens have been 
observed at this relatively high radio frequency implying quite high Rotation 
Measures (RMs) and strong regular magnetic fields. In the present paper we 
extend this work to a region closer to the Galactic centre, where Faraday 
effects are expected to be even more considerable.

Since the first detection of diffuse polarized emission from the Milky Way
Galaxy \citep{wsb+62,wsp62} it became evident that synchrotron emission is 
highly polarized. Radio polarization observations provide the most important 
data to study the magnetized warm interstellar medium (WIM), whose properties 
have not been fully understood yet. The separation of the thermal emission from 
the dominating non-thermal synchrotron component at low frequencies is also 
needed to determine the properties of the WIM. The inner Galactic plane is 
particularly interesting, but requires high-frequency observations to overcome 
strong depolarization effects. Radio continuum emission predominantly 
originates in spiral arms, and there are several arms along the line-of-sight 
towards the inner Galaxy \citep{hhs09}. 

Several total-intensity surveys covering the Galactic plane have been 
already conducted \citep[see the review by][]{wie05} and are accessible via 
the ``MPIfR Survey Sampler"\footnote{http://www.mpifr.de/survey.html}.
Among these surveys the Effelsberg $\lambda$11~cm survey \citep{rfrr90} and 
$\lambda$21~cm survey \citep{rrf90} include the inner Galaxy and have latitude 
extensions of $|b|\leq5\degr$ and $|b|\leq4\degr$, respectively. These surveys 
have high resolution and sensitivity, so that they are well suited to study 
large-scale diffuse emission. Early high-frequency surveys such as the Parkes 
$\lambda$6~cm survey \citep{hcs78} and the Nobeyama $\lambda$3~cm survey 
\citep{hsn+87} also cover the inner Galactic plane. However, their latitude 
extents are $|b|\leq2\degr$ and $|b|\leq1\fdg5$, respectively, too narrow to 
fully trace the layer of the WIM.

Polarization observations reveal the orientation of magnetic fields after 
correcting for Faraday rotation along the line-of-sight, which is produced by 
the magnetized diffuse WIM and magnetized thermal clumps. The latter are called 
Faraday Screens \citep{wr04}. Some \ion{H}{II} regions were identified as 
Faraday Screens, but often Faraday Screens do not have a counterpart in total 
intensity observations at centimetre bands. Faraday rotation also causes 
depolarization \citep{sbs+98}. The ``polarization horizon" was defined by 
\citet{luk02} as the distance beyond which entire depolarization occurs. In 
other words, the magnetized WIM is Faraday-thick beyond that distance 
\citep{sbs+98}. The polarization horizon depends on wavelength and increases
towards shorter wavelengths, where Faraday depolarization is less important 
\citep{sbs+98}. 

Previous polarization surveys of the Galactic plane have been reviewed by 
\citet{rei06}. Some were carried out with single-dish telescopes such as the 
Effelsberg $\lambda$11~cm survey \citep{jfr87,drrf99} and the Parkes 
$\lambda$13~cm survey \citep{dhjs97}, and others by interferometers such as the 
Canadian Galactic Plane Survey \citep[CGPS,][]{tgp+03,ulgk03,lrr+10} and the 
Southern Galactic plane survey \citep[SGPS,][]{gdm+01,hgm+06} both at 
$\lambda$21~cm. Except for the CGPS survey all the other surveys cover large 
parts of the inner Galaxy. The polarized emission observed by the Effelsberg 
telescope at $\lambda$11~cm and by the Parkes telescope at $\lambda$13~cm in 
the inner Galaxy probably originates in the Sagittarius--Carina arm 
\citep{drrf99}, which implies a polarization horizon of about 1~kpc at Galactic 
longitudes of about 30$\degr$. The polarized emission in the SGPS survey at 
$\lambda$21~cm was claimed to originate from the Crux arm \citep{gdm+01}, 
corresponding to a polarization horizon of about 3.5~kpc at longitudes of about 
330$\degr$. This is surprising, as the polarization horizon at $\lambda$21~cm 
is expected to be nearer than that at $\lambda$11~cm. However, these two surveys
have not included polarized large-scale structures, which is crucial for any 
scientific interpretation of polarized structures resulting from Faraday 
effects \citep{rei06}. 

The paper is organized as follows: We briefly describe the observations and 
data processing in Sect.~2. Survey maps as well as zero-level restorations of 
$U$ and $Q$ are presented in Sect.~3. Discrete objects and polarized structures 
are discussed in Sect.~4 and Sect.~5, respectively. The separation of thermal 
and non-thermal emission components is conducted in Sect.~6. Sect.~7 gives a 
summary. The survey data will be available from the ``MPIfR Survey Sampler"
and the web-page of the $\lambda$6~cm survey at 
NAOC\footnote{http://zmtt.bao.ac.cn/6cm/} after completion of the project.

\section{Observations and data processing}

\subsection{General}

We have carried out the Sino-German $\lambda$6~cm polarization survey of the 
Galactic plane by using the Urumqi 25~m telescope of the NAOC. The results of 
the first small survey section covering $122\degr\leq l\leq129\degr$ and 
$|b|\leq5\degr$ were published in Paper~I. In Paper~II results for the region 
of $129\degr\leq l\leq230\degr$ were presented. The region 
$60\degr\leq l\leq122\degr$ will be discussed in a forthcoming paper. In Papers 
I and II numerous Faraday Screens without total intensity correspondence 
hosting strong regular magnetic fields were the most outstanding polarization 
features. In this paper we present the results for the inner Galaxy in the 
range of $10\degr\leq l\leq60\degr$ and $-5\degr\leq b\leq 5\degr$, where 
diffuse emission is the most prominent.

To carry out the survey, the Urumqi telescope was equipped with a $\lambda$6~cm 
receiving system constructed at the MPIfR. The $\lambda$6~cm survey has a 
resolution of $9\farcm5$ and high sensitivity to match the available 
lower-frequency surveys from the Effelsberg telescope. The system temperature 
was about 22~K towards the zenith. The central frequency was 4.8~GHz 
($\lambda=6.25$~cm used for all calculations throughout the paper), and the 
original bandwidth was 600~MHz. To avoid the influence from stationary 
communication satellites, the bandwidth was often reduced to 295~MHz. The 
corresponding central frequency then is 4.963~GHz. The system gain is 
$T_{\rm B}[{\rm K}]/S[{\rm Jy}]=0.164$. More detailed information about the 
receiving system has already been presented in Paper~I.

The Galactic plane was scanned in both longitude and latitude directions. For 
the longitude scans, the plane was divided into sections with a typical 
size of $7\degr\times2\fdg6$. The size varied to include large structures 
completely and avoid strong sources at the edge areas. For the latitude scans, 
the plane was split into sections of $2\degr\times10\degr$ in size. Two 
neighbouring sections overlapped by $0\fdg1$ to ease the later adjustment of 
intensity levels. The separation of sub-scans was $3\arcmin$ to assure full 
sampling. The scanning velocity was $4\degr/{\rm min}$, thus a section was 
finished after about 100~minutes. Therefore observations did not span too large 
ranges in azimuth and elevation, and the influence of varying ground emission 
was significantly reduced. Observations were always made at night time with 
clear sky. 

The primary calibrator was 3C~286 with a flux density of 7.5~Jy, a polarization 
percentage of 11.3\% and a polarization angle of 33$\degr$. 3C~48 and 3C~138 
served as secondary calibrators. The unpolarized calibrators were 3C~295 and 
3C~147. A calibration source was observed before and after each survey section. 

The dual-channel receiving system provided left-hand ($LHC$) and right hand 
($RHC$) polarization channels, and the correlation of the two channels provided 
Stokes $U$ and $Q$. The average of the $LHC$ and $RHC$ channels provided $I$, 
assuming no circular polarization. Raw data were recorded on disk. For each 
individual sub-scan a linear-fit based on the two ends was subtracted, $U$ and 
$Q$ channels were corrected for phase and parallactic angle, and finally raw 
maps were calculated. The raw maps of $I$, $U$ and $Q$ were stored in 
\textsc{NOD2}-format \citep{has74}. Further processing was based on a standard 
procedure developed for Effelsberg continuum and polarization observations as 
detailed in Paper~I and Paper~II. Briefly, spikes were removed, baselines were 
further corrected, and scanning effects were suppressed by the ``unsharp 
masking" method \citep{sr79}. Each processed map was multiplied by a factor 
determined from the calibration source data to convert map units into a physical
 scale. The baselines of longitude scans were adjusted in respect to the level 
of the latitude scans. Therefore structures exceeding about $10\degr$ in 
latitude direction were missing in the original survey maps. Afterwards maps 
observed in these orthogonal directions were added in the Fourier domain with 
the spatial weights defined by the typical scale of the scanning noise 
\citep{eg88}. To eliminate position errors and shifts among the maps, we picked 
up strong sources and compared them with much more precise source positions 
from the NVSS survey \citep{ccg+98} and made corrections if necessary. From the 
processing it is clear that the $I$, $U$ and $Q$ maps have the latitude ends 
set to zero. The polarization angle ($\psi$) and polarized intensity ($PI$) 
were calculated as $\psi=\frac{1}{2}{\rm atan}\frac{U}{Q}$ and 
$PI=\sqrt{U^2+Q^2-1.2\sigma_{U,\,Q}^2}$ \citep{wk74}, where $\sigma_{U,\,Q}$ is 
the average rms-noise of $U$ and $Q$. Note that polarization angles are 
relative to the Galactic north and run anti-clockwise.
   
\begin{figure}[!htbp]
\centering
\includegraphics[width=6.5cm,angle=-90]{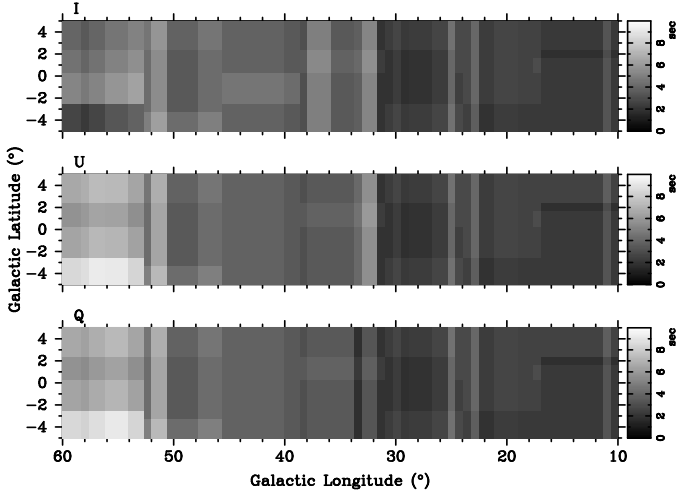}
\caption{Effective integration time in sec for $I$, $U$ and $Q$ from {\it top} 
to {\it bottom}. The time was scaled by a factor of 0.5 in case the 
observations were made with a reduced bandwidth of $\Delta\nu=295$~MHz. }
\label{intT}
\end{figure}

The rms-noise for the surveyed area is about 1~mK~$T_{\rm B}$ for $I$, and 
0.5~mK~$T_{\rm B}$ for the $U$, $Q$ and $PI$ maps, which was measured from 
nearly ``empty" regions without obvious emission features or gradients. The 
theoretical rms-noise for $I$, $U$ and $Q$ maps assuming a perfectly stable 
receiving system is calculated from the system temperature $T_{\rm sys}$, the 
bandwidth $\Delta\nu$, and the integration time $\Delta t$ (see Fig.~\ref{intT})
 according to $\sigma\propto T_{\rm sys}/\sqrt{\Delta\nu\Delta t}$. Typically 
four latitude and three longitude coverages were observed for one section. With 
a bandwidth of 295~MHz and an integration time of 0.75~s for one coverage, 
the theoretical rms-noise is 0.8~mK~$T_{\rm B}$ for $I$ and 0.6~mK~$T_{\rm B}$ 
for $U$ and $Q$. Thus the measured rms-noise is close to the expectations for a
perfect receiving system. The rms-noise varies because of different bandwidths 
used and effective integration time (see Fig.~\ref{intT}). Generally the 
rms-noise becomes larger towards lower longitude regions, where the diffuse 
emission is stronger and the typical elevations are lower. The systematic 
uncertainty of the polarization angle is about $1\degr$, and the calibration 
scale error is about 5\%. Both estimates were inferred from observations of 
3C~286.  

\subsection{Instrumental polarization}

Instrumental effects related to the properties of the Urumqi telescope, the 
$\lambda$6~cm feed and receiving system cause polarization from unpolarized 
sources. Thus intrinsically unpolarized \ion{H}{II}~regions show polarized 
intensity, which exhibits a ring-like pattern for compact sources. The radius 
of the ring is about that of the beam width and the maximum intensity is 
measured to be less than 3\% of the peak value of total intensity 
\citep{srh+06}. Instrumental polarization at this level is not important 
towards the anti-centre region, where just a few strong sources are located. In 
the inner Galaxy, however, there are many strong compact \ion{H}{II} regions, 
which together add up so that their instrumental polarization becomes 
significant and a proper cleaning is required.

\begin{figure}[!htbp]
\centering
\includegraphics[angle=-90,width=8cm]{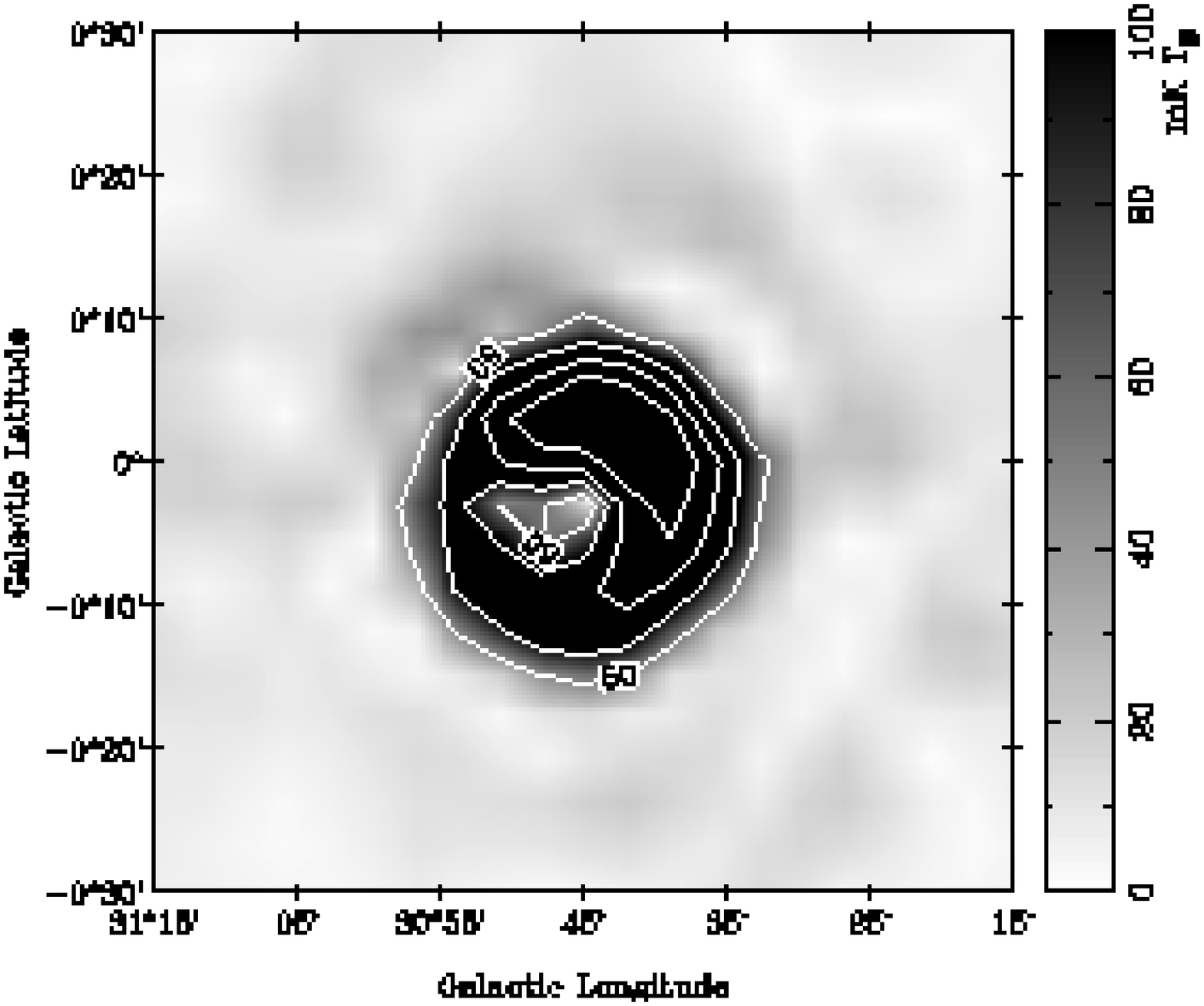}
\includegraphics[angle=-90,width=8cm]{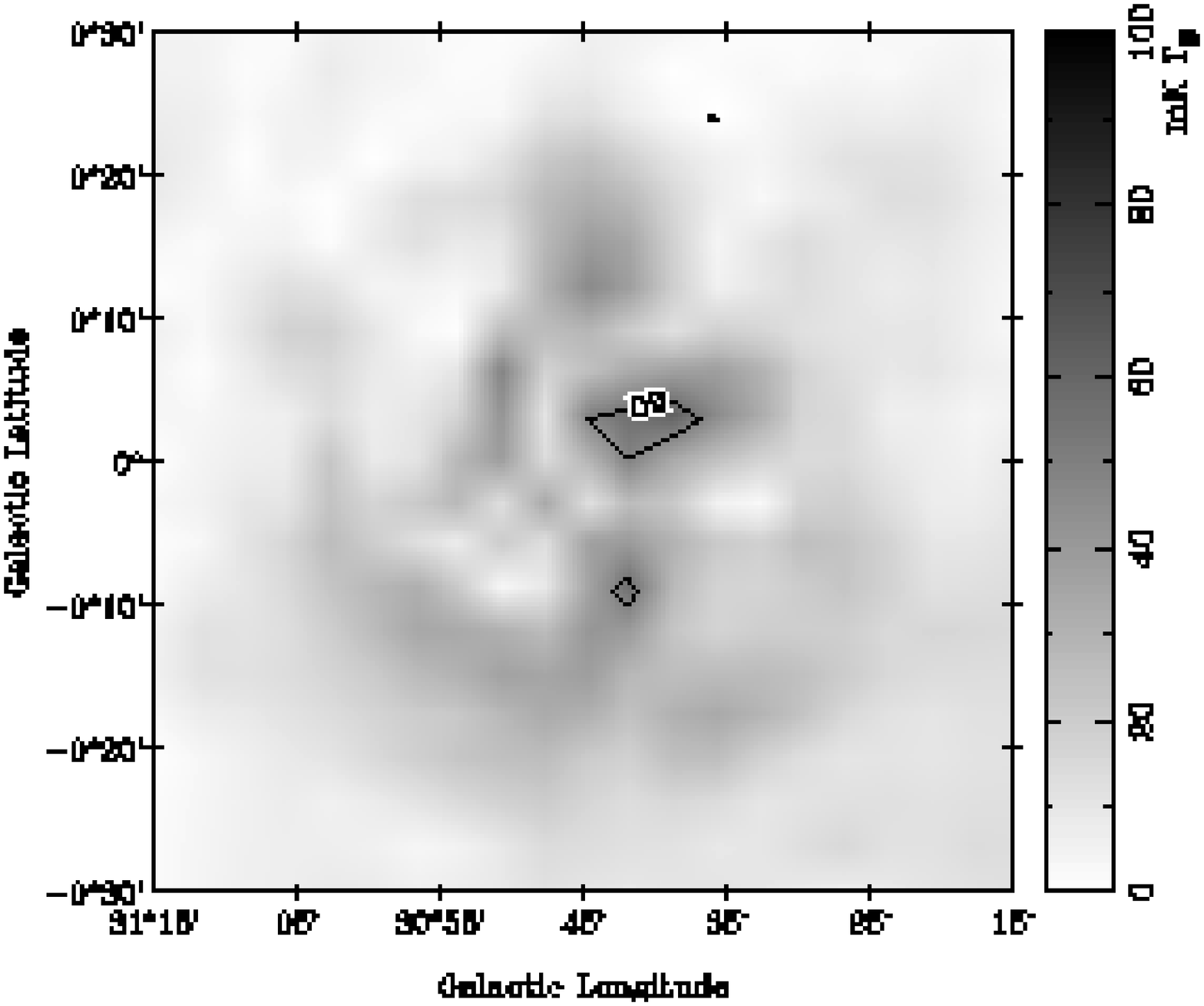}
\caption{Example of instrumental polarization cleaning applied to the 
$\lambda$6~cm survey polarized intensity data for the case of the \ion{H}{II} 
region G30.7$-$0.1. The {\it upper panel} shows the original polarized 
intensity and the {\it lower panel} the cleaned map. The contours start at 
50~mK~$T_{\rm B}$ and run in steps of 50~mK~$T_{\rm B}$.}
\label{clean}
\end{figure}

The cleaning method we applied has already been described in Paper~I in some 
detail. By observing unpolarized calibrators we obtained maps of total 
intensity ($I_{\rm cal}$) and polarization ($U_{\rm cal}$ and $Q_{\rm cal}$). 
Here the polarization is assumed to be completely from instrumental impurity, 
which can be quantitatively represented as the convolution between the total 
intensity of the calibrator and the instrumental terms ($U_{\rm inst}$ and 
$Q_{\rm inst}$) as $U_{\rm cal}=I_{\rm cal}\otimes U_{\rm inst}$ and 
$Q_{\rm cal}=I_{\rm cal}\otimes Q_{\rm inst}$. We performed a de-convolution 
in the Fourier domain to derive the instrumental terms. We convolved the survey 
total intensity maps with the instrumental polarization terms to determine the 
instrumental polarization, and subtracted these instrumental contributions from 
the observed polarization maps to obtain the cleaned maps. This method was 
already successfully applied earlier by \citet{sris87}.
 
We used the \ion{H}{II} region RCW~174 ($l=28\fdg8$, $b=3\fdg5$) as calibrator,
which shows a high signal-to-noise ratio. It was used to clean the survey maps 
fairly well. In Fig.~\ref{clean} we show the elimination of the instrumental 
polarization for the \ion{H}{II} region G30.7$-$0.1. The total intensity of 
this \ion{H}{II} region is about 11~K~T$_{\rm B}$, and the polarized intensity 
caused by instrumental effects is about 0.3~K~T$_{\rm B}$, which is reduced to 
about 0.06~K~T$_{\rm B}$, or about 0.5\%, after cleaning.  

Generally the residual instrumental polarization is well below 1\% for the 
cleaned maps. Sources with total intensities larger than 250~mK~$T_{\rm B}$ 
have a residual polarization at about the 5$\sigma$ level. Compared to the 
large fluctuations of diffuse polarized intensities along the Galactic plane 
these residual instrumental effects are usually negligible.

\section{Survey results}

As described above $I$, $U$ and $Q$ intensities were set to zero at both 
latitude ends ($b=\pm5\degr$), because of baseline fitting to remove the ground 
radiation contributions. This implies that the maps have a relative zero-level.
For $I$ maps this is equivalent to an intensity offset, though the observed 
small-scale structures are not influenced. Thus zero-level setting for total 
intensity is only crucial when the diffuse emission is quantitatively assessed 
as we do below to separate thermal and non-thermal emission (see Sect.~6).
However, polarized intensities and polarization angles depend on $U$ and $Q$ 
non-linearly, so that their morphology and structures were significantly 
changed due to the missing large-scale emission \citep[e.g. ][]{rei06}. It is 
thus essential to recover it for a correct physical interpretation. 

\subsection{Zero-level restoration for $U$ and $Q$}\label{zero_rest}

Unfortunately no $\lambda$6~cm polarization data exist at an absolute 
zero-level, which we might refer to. The on-going C-Band All-Sky Survey 
(CBASS)\footnote{http://www.astro.caltech.edu/cbass} is expected to provide
the missing large-scale emission. The CBASS has a resolution of $44\arcmin$ and 
a sensitivity of about 0.1~mK, which is very similar to that of the Urumqi
$\lambda$6~cm polarization survey when smoothed to the same angular resolution. 
Thus the CBASS data are well suited for zero-level restoration. However, as 
shown later, the required absolute zero-level of polarization at $\lambda$6~cm 
must be accurate to about 1~mK to be superior to the zero-level restoration 
scheme we present in the following.

We developed a calibration scheme for the $\lambda$6~cm survey in Paper~I, 
where the correct zero-levels of $U$ and $Q$ were tied to the three-year 
WMAP K-band (22.8~GHz) data \citep{phk+06}. Now we used the more precise 
five-year WMAP results \citep{hwh+09}. The K-band polarization data include all 
large-scale structures. We calculated the corresponding polarized intensities 
at 4.8~GHz including all large-scale components ($PI\rm^{4.8\,GHz}_{zero}$) by: 
\begin{equation}\label{abso_pi}
PI{\rm^{4.8\,GHz}_{zero}}=PI{\rm^{22.8\,GHz}}
\left(\frac{4.8}{22.8}\right)^{\beta_{PI}},
\end{equation}
and subsequently for $U$ and $Q$: 
\begin{equation}\label{abso_uq}
\begin{array}{rcl}
U\rm ^{4.8\,GHz}_{zero}&=&PI\rm^{4.8\,GHz}_{zero}
\sin\left(2\psi^{22.8\,GHz}+\mathcal{R}\lambda^2\right)\\[5mm]
Q\rm^{4.8\,GHz}_{zero}&=&PI\rm^{4.8\,GHz}_{zero}
\cos\left(2\psi^{22.8\,GHz}+\mathcal{R}\lambda^2\right).
\end{array}
\end{equation}
$\psi$ is the polarization angle, and $\beta_{PI}$ is the spectral index for 
the polarized intensity ($T\propto\nu^\beta$ with $T$ being the brightness 
temperature and $\nu$ the frequency) and $\mathcal{R}$ is the Faraday depth. 
Assuming the interstellar medium to be a uniform slab, the Faraday depth and RM 
observed from diffuse polarized emission are related 
$\mathcal{R}=2\rm RM$ \citep{sbs+98}. For extragalactic sources and pulsars the 
Faraday depth is equivalent to RM.

As already described in Paper~I, we convolved the $U$ and $Q$ maps of both the 
$\lambda$6~cm survey and the WMAP K-band survey to a resolution of $2\degr$, 
calculated the intensities according to Eqs.~(\ref{abso_pi}) and 
(\ref{abso_uq}), estimated the differences to the observed $\lambda$6~cm maps, 
and finally added the differences to the original $U$ and $Q$ $\lambda$6~cm 
maps. However, towards the inner Galaxy significant depolarization might occur 
even at $\lambda$6~cm. The observed polarized $\lambda$6~cm emission originates 
at shorter distances than that observed at K-band. In this case, the above 
scheme needs to be modified, otherwise too much polarization will be added. We 
therefore need to determine the spectral index of the polarized intensity, 
investigate the polarization horizons at both wavelengths, and check the 
Faraday depths before applying any zero-level restoration.  

\subsubsection{Spectral index of polarized intensity}\label{spmodel}

Diffuse polarized emission originates from Galactic synchrotron radiation, 
while total intensities are a mixture of synchrotron and free-free emission.
Therefore the total intensity spectral index is not that of the synchrotron 
emission. Polarization observations at low frequencies suffer from significant 
depolarization, and are thus not suited to determine the synchrotron spectral 
index. At high frequencies, polarized intensities are usually very weak because 
of the generally steep spectrum. Fortunately in the inner Galaxy polarized 
emission at both the WMAP K- and Ka-band (33~GHz) \citep{hwh+09} is strong 
enough for this purpose. 

\begin{figure}[!htbp]
\centering
\resizebox{0.46\textwidth}{!}{\includegraphics[angle=-90]{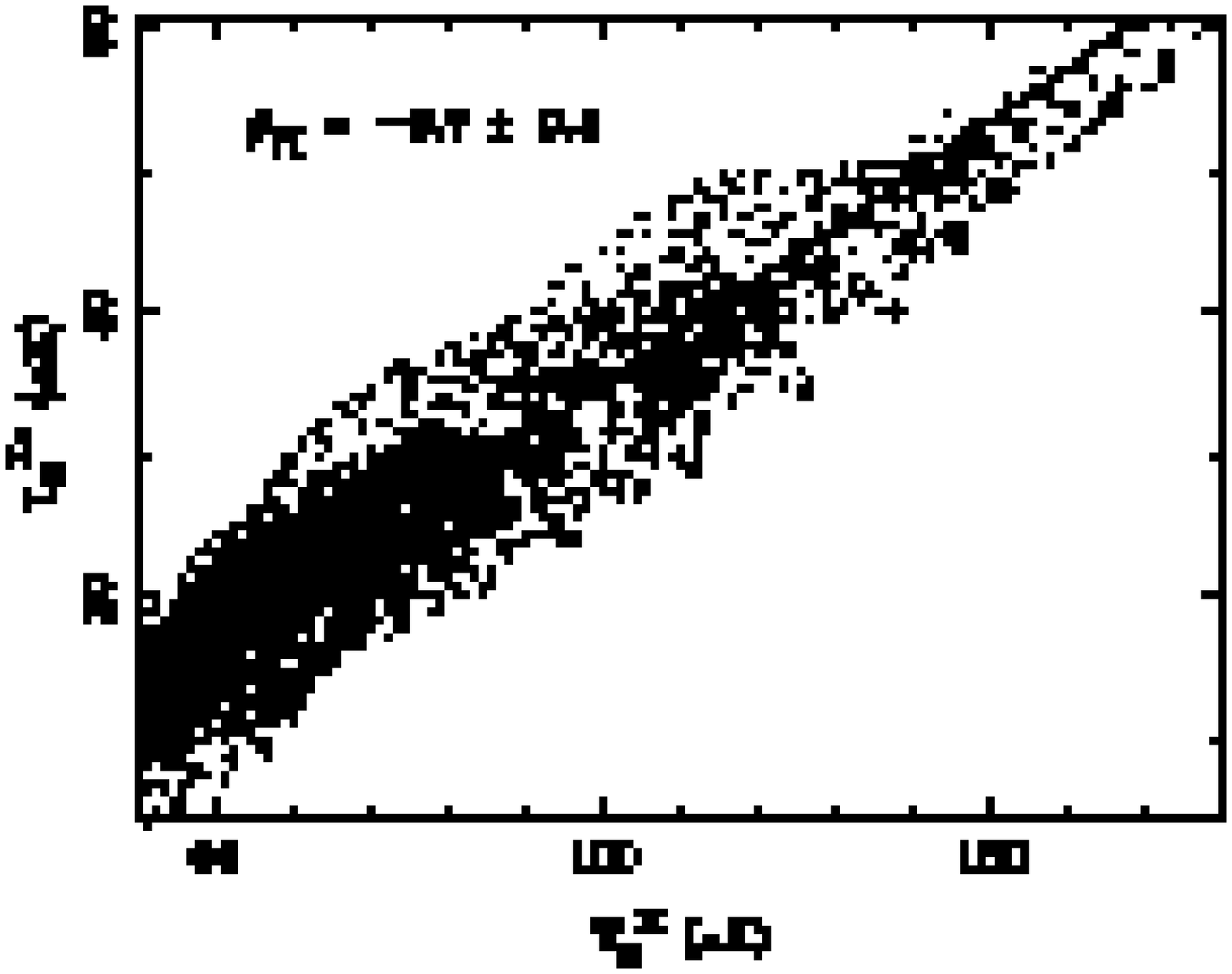}}
\resizebox{0.46\textwidth}{!}{\includegraphics[angle=-90]{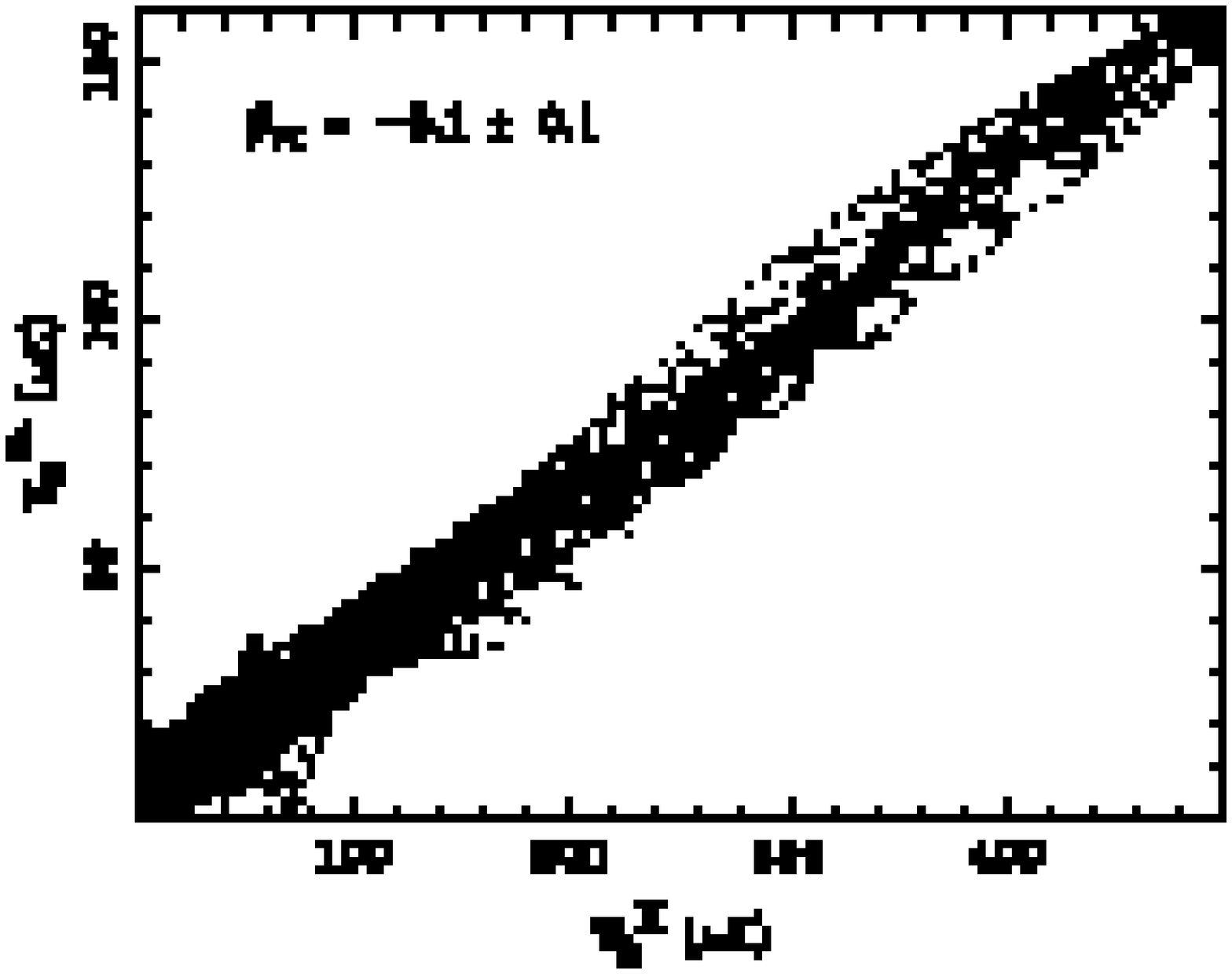}}
\caption{TT-plot between WMAP K-band (22.8~GHz) and Ka-band (33~GHz) for 
polarized intensities for the region of $40\degr\leq l\leq60\degr$ ({\it upper 
panel}) and $10\degr\leq l\leq30\degr$ ({\it lower panel}) within the latitude 
range of $|b|\leq5\degr$.}
\label{beta_kka}
\end{figure}

We performed TT-plots \citep{tpkp62} for polarized intensity between K- and 
Ka-band for every $10\degr\times10\degr$ section for the region 
 $10\degr<l<60\degr$ and $|b|<5\degr$, and found a trend of steepening of 
spectral index from larger to smaller longitudes. The transition region is 
between longitudes $30\degr$ and $40\degr$. For the same area \citet{rr88a} 
reported a spectral steeping in total intensity between 408~MHz and 1420~MHz. 
We modelled the spectral index as follows: For the region 
$10\degr\leq l\leq30\degr$, the spectral index is $\beta_{PI}=-3.1$, for the 
region  $40\degr\leq l\leq60\degr$, the spectral index is $\beta_{PI}=-2.7$ 
(Fig.~\ref{beta_kka}), and for the transition region the spectral index is a 
linear interpolation between $-2.7$ and $-3.1$. Note that the spectral index 
error for the region $10\degr\leq l\leq30\degr$ is $\Delta\beta = 0.1$ from the 
TT-plot, but increased to $\Delta\beta = 0.4$ for the area 
$40\degr\leq l\leq60\degr$ caused by a lower signal-to-noise ratio there. 

At K- and Ka-bands, depolarization is negligible and the polarization from 
spinning dust can be neglected as well \citep{gow+10}. This means that 
polarized emission at both bands shows intrinsic synchrotron emission. The 
spectral index derived from polarization data applies also for the Galactic
total intensity synchrotron emission component, i.e. 
$\beta_{\rm syn}=\beta_{PI}$.

\subsubsection{Polarization horizon} \label{dph_sim}

\begin{figure}[!htbp]
\centering
\includegraphics[angle=-90,width=9cm]{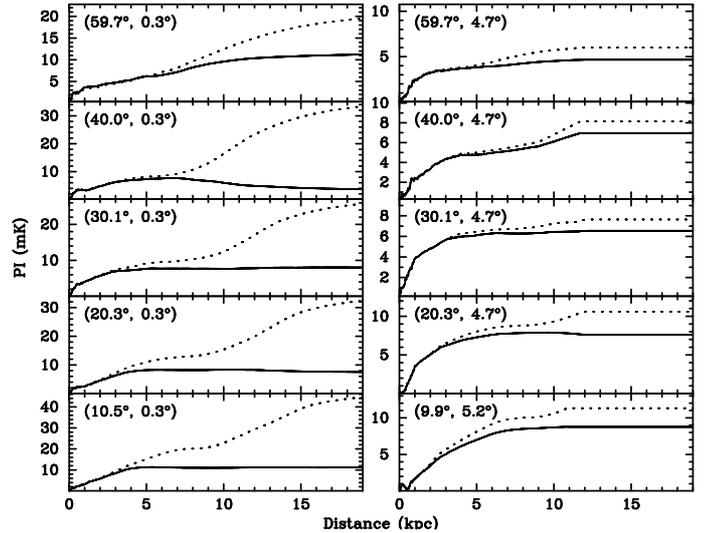}
\caption{Simulated polarized intensity at $\lambda$6~cm (4.8~GHz) and K-band 
(22.8~GHz) versus distance for various directions as indicated by the Galactic 
coordinates in each panel. The dotted lines show the polarized intensity at 
K-band scaled to $\lambda$6~cm with a spectral index as modelled in 
Sect.~\ref{spmodel}. The solid lines show $PI$ at $\lambda$6~cm.}
\label{dph}
\end{figure}

To check whether the $\lambda$6~cm and K-band surveys observe polarized 
emission from the same polarization horizon, we performed simulations to show 
polarized intensities at these two frequencies versus distances for several 
directions along the Galactic plane. The simulations were based on the 
\textsc{hammurabi} code developed by \citet{wjr+09} and the new Galactic 
3D-emission model developed by \citet{srwe08} using the modifications for 
high-resolution simulations for selected patches by \citet{sr09}, where a 
Kolmogorov spectrum for the random magnetic fields was included. For each 
line of sight, the complex polarization $\mathcal{P}$ was calculated following 
\citet{srwe08} as, 
\begin{equation} 
\displaystyle{
\mathcal{P}=\int_0^{r_{\rm max}}P(r)\exp\left(
2\mathbf{i}(\psi_0(r)+{\rm RM}(r)\lambda^2)\right){\rm d}r}, 
\end{equation}
where $P(r)$ is the polarized intensity at $r$, $\psi_0(r)$ is the intrinsic 
polarization angle, and the integral was conducted from the observer to a 
specified distance of $r_{\rm max}$ along the line of sight. The polarized 
intensity is the absolute value of $\mathcal{P}$. The positions 
were selected to be in the plane and above the plane at latitudes of about 
$5\degr$ and longitudes between $10\degr$ and $60\degr$. The simulations have 
an angular resolution of about $30\arcsec$. The results are shown in 
Fig.~\ref{dph}, where the data were averaged for an area with a radius of 
$0\fdg1$ centred at the coordinates marked in each panel. As explained 
by \citet{sr09} the coordinates of the target patches for simulations had to be 
adapted to the HEALPix \citep{ghb+05} projection scheme. The polarized 
intensities at 22.8~GHz were scaled to 4.8~GHz with a spectral index detailed 
in Sect.~\ref{spmodel} to facilitate the comparison. 

Figure~\ref{dph} shows that at $\lambda$6~cm the polarization horizon in the 
Galactic plane is about 3~kpc for $10\degr\leq l\leq30\degr$ and increases to 
about 5~kpc for $40\degr\leq l\leq60\degr$. Below we adopt 4~kpc as the 
$\lambda$6~cm polarization horizon for discussion. Figure~\ref{dph} also shows 
that about 20\%--30\% of the intrinsic Galactic polarized emission is observed 
in the Galactic plane at $\lambda$6~cm. Above the Galactic plane at latitudes 
of about $5\degr$, only less than about 20\% of polarized emission is missed 
when compared to the scaled K-band data. At the latitude edges of the 
$\lambda$6~cm survey the polarized emission traced at both wavelengths 
originate from volumes which are not significantly different.

\subsubsection{Faraday depth} 
  
\begin{figure}[!htbp]
\centering
\includegraphics[width=5cm,angle=-90]{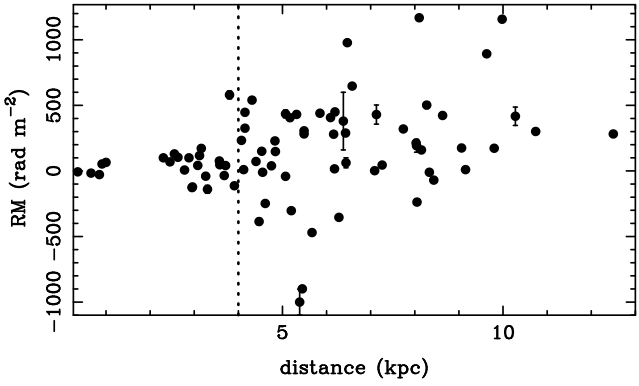}
\caption{Pulsar RMs versus distance based on the NE2001 electron density model.
For most of the data the RM error bars are too small to be seen.}
\label{psrrm}
\end{figure}

The Faraday depth in a certain direction can be estimated from RMs of 
extragalactic sources, pulsars and diffuse emission. RMs of extragalactic 
sources contain all contributions along the line-of-sight from the Sun to the 
Galactic outskirts. However, the polarization horizon inferred from the 
simulations is about 4~kpc at $\lambda$6~cm, which is shorter than the 
path length across the Galaxy. Therefore RMs of extragalactic sources cannot be 
used directly. We retrieved RMs of pulsars in the region of $10\degr<l<60\degr$ 
and $|b|<5\degr$ from the ATNF pulsar catalogue~\citep{mhth05}. The distances 
of these pulsars were estimated by using the NE2001 Galactic electron density 
model \citep{cl02}. Pulsar RMs versus distances are displayed in 
Fig.~\ref{psrrm}. The average Faraday depth at distances up to the 
$\lambda$6~cm polarization horizon of 4~kpc was estimated from the standard 
deviation of pulsar RMs (excluding one RM outlier of 580~rad~m$^{-2}$) to be 
about 22~rad~m$^{-2}$. This corresponds to an angle rotation of about $5\degr$ 
at $\lambda$6~cm and we conclude that the Faraday depth in Eq.~(\ref{abso_uq}) 
can be neglected.   

Variations of the Faraday depth with longitude cannot be reliably obtained from
the small number of pulsar RMs shown in Fig.~\ref{psrrm}. We simulated $U$ and 
$Q$ maps at 4.8~GHz and 22.8~GHz using the Galactic 3D-emission models by 
\citet{srwe08}, derived polarization angle maps at both frequencies, and 
calculated the RM map. RMs are found to be within $\pm10$~rad~m$^{-2}$ from 
$10\degr$ longitude to $50\degr$ longitude, which is consistent with the result 
obtained from pulsar RMs. From $l=50\degr$ to $l=57\degr$ the RMs can be as 
large as about 50~rad~m$^{-2}$ corresponding to an angle rotation of about 
$11\degr$, which in principle should be accounted for. As shown below, the 
zero-level correction to be added to $U$ and $Q$ for this region is less than  
1~mK~$T_{\rm B}$, which is about twice the rms-noise. Therefore we neglect this 
RM correction.

\subsubsection{A modified scheme for $U$ and $Q$ zero-level restoration}

Because of the different polarization horizons at $\lambda$6~cm and at K-band 
in the Galactic plane, the scaled difference to the K-band polarized emission 
cannot be used for a correction of the polarization zero-levels. We thus 
modified the correction scheme of Paper~I by referring to the high-latitude 
regions between $4\fdg5\leq|b|\leq5\degr$, where the $\lambda$6~cm and K-band 
surveys observe polarized emission from nearly the same volume. A similar 
modification was made in Paper~II, where the polarized emission in the plane is 
very weak.

\begin{figure}[!htbp]
\centering
\includegraphics[angle=-90,width=8cm]{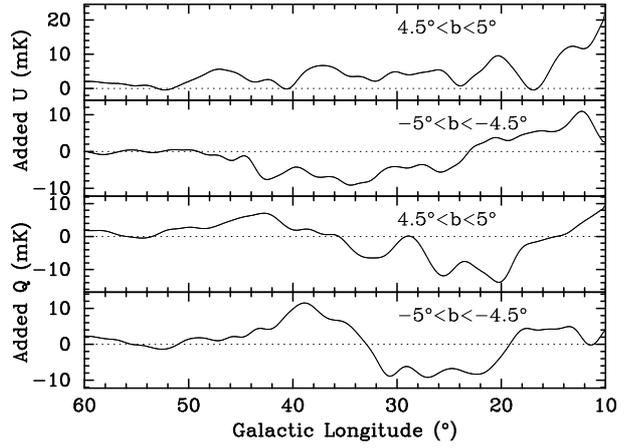}
\caption{Longitude profiles showing $U$ and $Q$ corrections
averaged for $4\fdg5\leq|b|\leq5\degr$ (see Sect. 3.1.4. for details).}  
\label{uqoffset}
\end{figure}

We obtained the differences between the $\lambda$6~cm and K-band $U$ and $Q$ 
maps for the two high-latitude regions. The correction values were calculated 
by linear interpolation along latitude and added to the original $U$ and $Q$ 
data. The $U$ and $Q$ corrections averaged from the difference maps for 
$4\fdg5\leq|b|\leq5\degr$ are plotted versus longitude in Fig.~\ref{uqoffset}. 
The values shown in Fig.~\ref{uqoffset} are the maximal and minimal corrections 
for $U$ and $Q$ being applied for each latitude scan. For the region of 
$50\degr\leq l\leq60\degr$ the average correction is about 0.6~mK~$T_{\rm B}$ 
for $U$ and 0.8~mK~$T_{\rm B}$ for $Q$, which justifies to neglect a Faraday 
depth dependent correction in view of the rms-noise of 0.5~mK~$T_{\rm B}$.

\begin{figure}[!htbp]
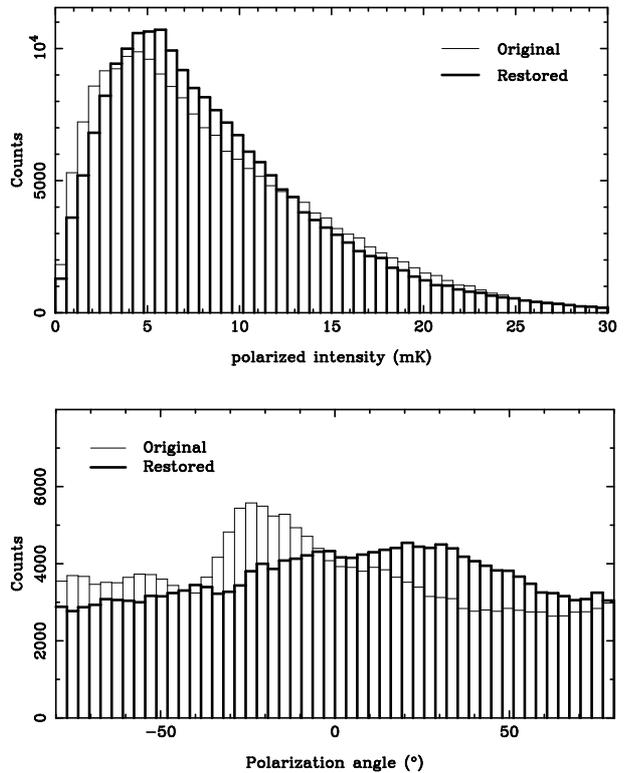

\centering
\includegraphics[angle=-90,width=8cm]{pi_hist.ps}
\\[5mm]
\includegraphics[angle=-90,width=8cm]{pa_hist.ps}
\caption{Pixel distribution of polarized intensities ({\it top}) and 
polarization angles ({\it bottom}) before and after baseline restoration.}
\label{pipa_dist}
\end{figure}

In Fig.~\ref{pipa_dist} we show the pixel distributions of polarized 
intensities and polarization angles. The distributions with and without adding 
the large-scale components are quite similar. This result largely differs from 
that obtained for the survey regions of the outer Galaxy as presented in 
Papers I and II, where the large-scale corrected polarization angle 
distribution peaks at $0\degr$.

\subsubsection{Accuracy of the zero-level restoration scheme}

The spectral index of polarized intensity has an uncertainty of 
$\Delta\beta = 0.1$ for the inner region $10\degr<l<30\degr$. This corresponds 
to a restoration uncertainty of less than 20\% or about 2~mK~$T_{\rm B}$ for 
areas where the $U$ and $Q$ values are as high as 10~mK~$T_{\rm B}$ 
(Fig.~\ref{uqoffset}). For the outer region $40\degr<l<60\degr$, the spectral 
index has a large uncertainty of $\Delta\beta = 0.4$, but the typical added $U$ 
and $Q$ values are around 1~mK~$T_{\rm B}$ (Fig.~\ref{uqoffset}). This again 
means an uncertainty of about 2~mK~$T_{\rm B}$ for the zero-level. To 
summarize: the uncertainty of the applied zero-level restoration for the entire 
survey region caused by the spectral index uncertainty used in the 
extrapolation from WMAP frequency towards 4.8~GHz is up to the $4\times\sigma$ 
level of the observed polarized intensity, but mostly below.

Based on the simulations presented in Sect.~\ref{dph_sim}, the $\lambda$6~cm 
survey could miss up to about 20\% of polarized intensity at the latitude 
edges. This is equivalent to a spectral index uncertainty of about 
$\Delta\beta = 0.1$. As already shown above, this does not have a significant 
influence on the results.

When the CBASS $\lambda$6~cm polarization survey becomes available, missing 
large-scale components of this survey can be added without extrapolation over a 
wide frequency range. Compared to the present extrapolation method, an 
improvement will be obtained in case the zero-level in $U$ and $Q$ can be 
measured with an accuracy better than 2~mK.

\subsection{Overview of the survey maps}

\begin{figure*}[!htbp]
\centering
\resizebox{0.95\textwidth}{!}{\includegraphics{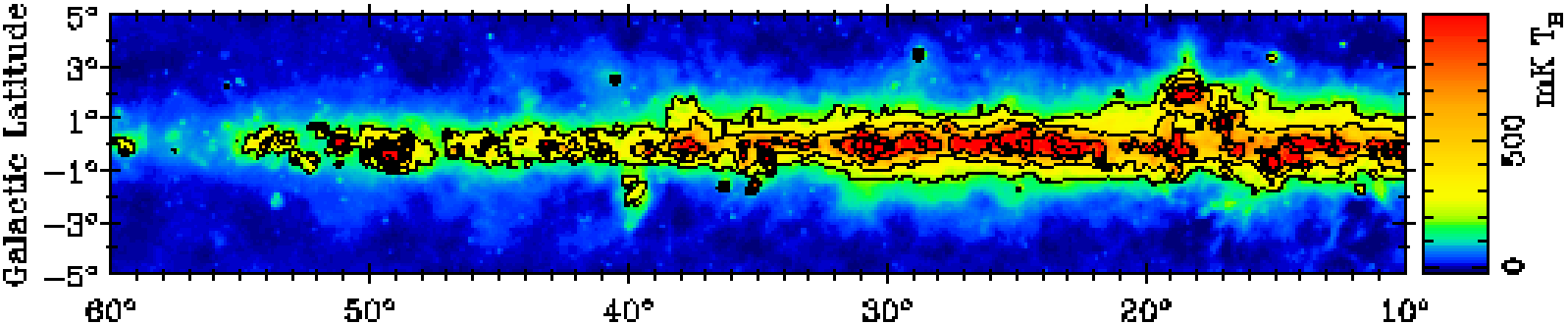}}
\resizebox{0.95\textwidth}{!}{\includegraphics{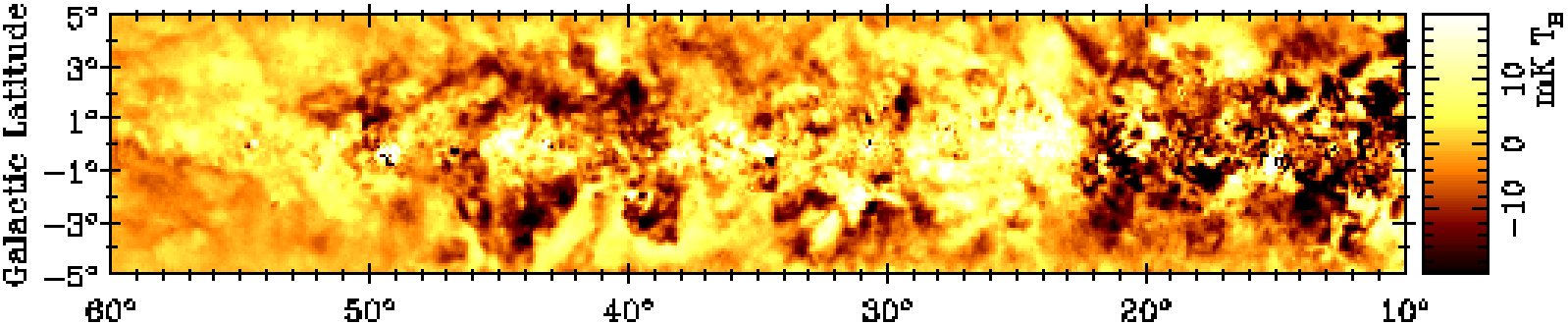}}
\resizebox{0.95\textwidth}{!}{\includegraphics{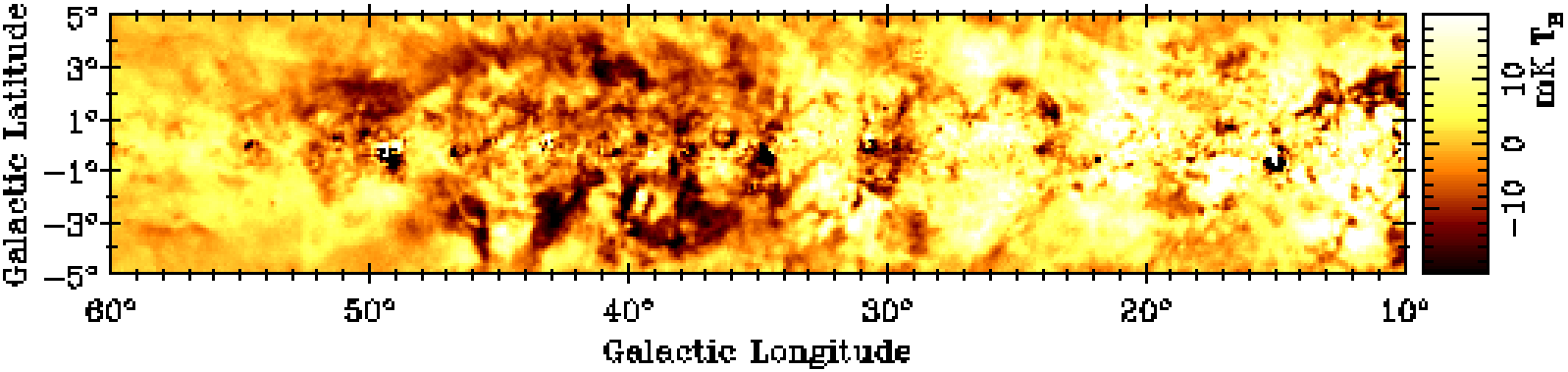}}
\caption{Maps at $\lambda$6~cm for the region $10\degr\leq l\leq60\degr$.  
From {\it top} to {\it bottom} $I$, $U$ and $Q$ maps are shown as observed. 
Overlaid contours on the $I$ map run in steps of $2^n\times200$~mK~$T_{\rm B}$ 
with $n=0,\,1,\,2,\,\ldots$.}
\label{g35all}
\end{figure*}

\begin{figure*}[!htbp]
\centering
\resizebox{0.95\textwidth}{!}{\includegraphics[angle=-90]{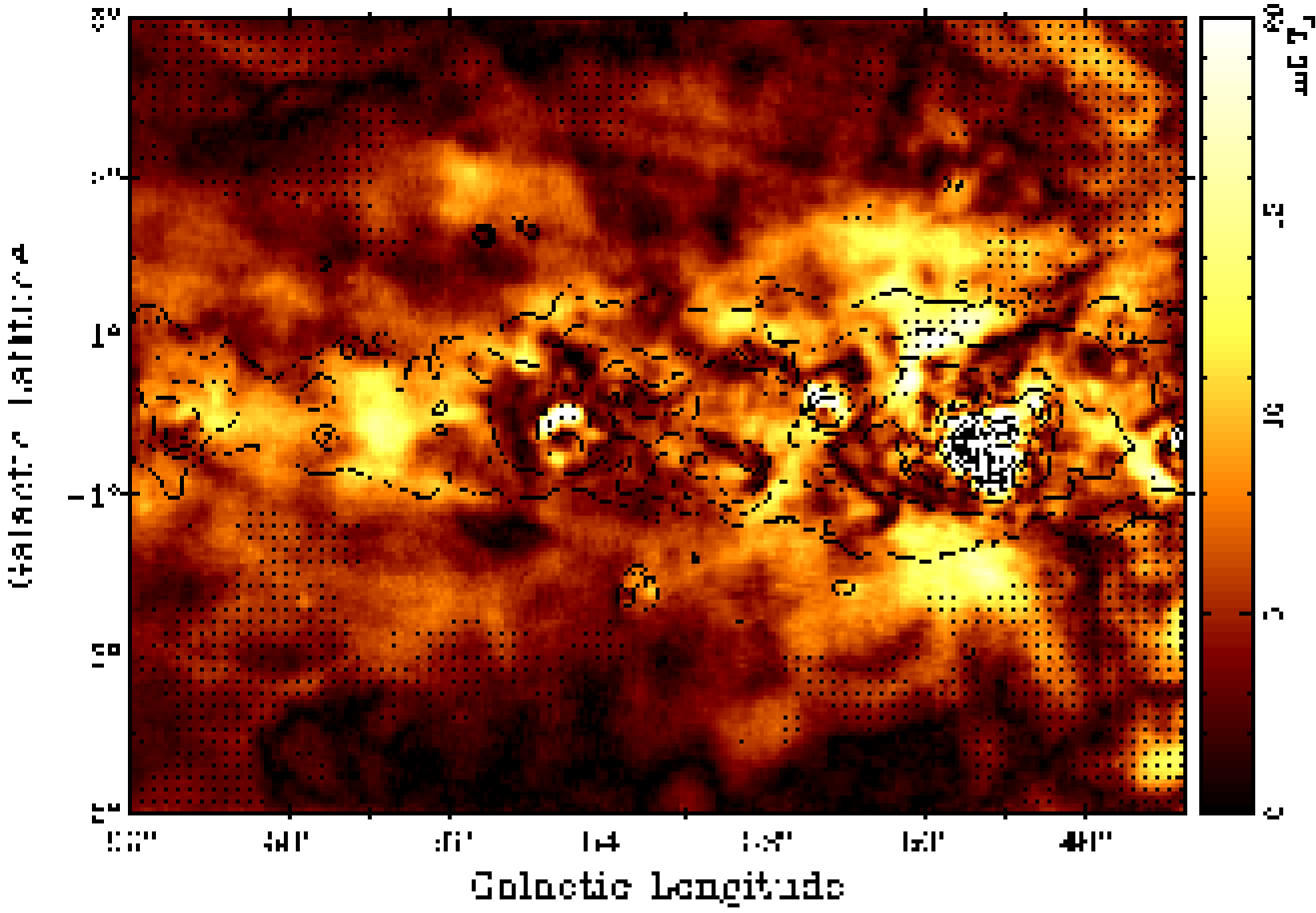}}
\resizebox{0.95\textwidth}{!}{\includegraphics[angle=-90]{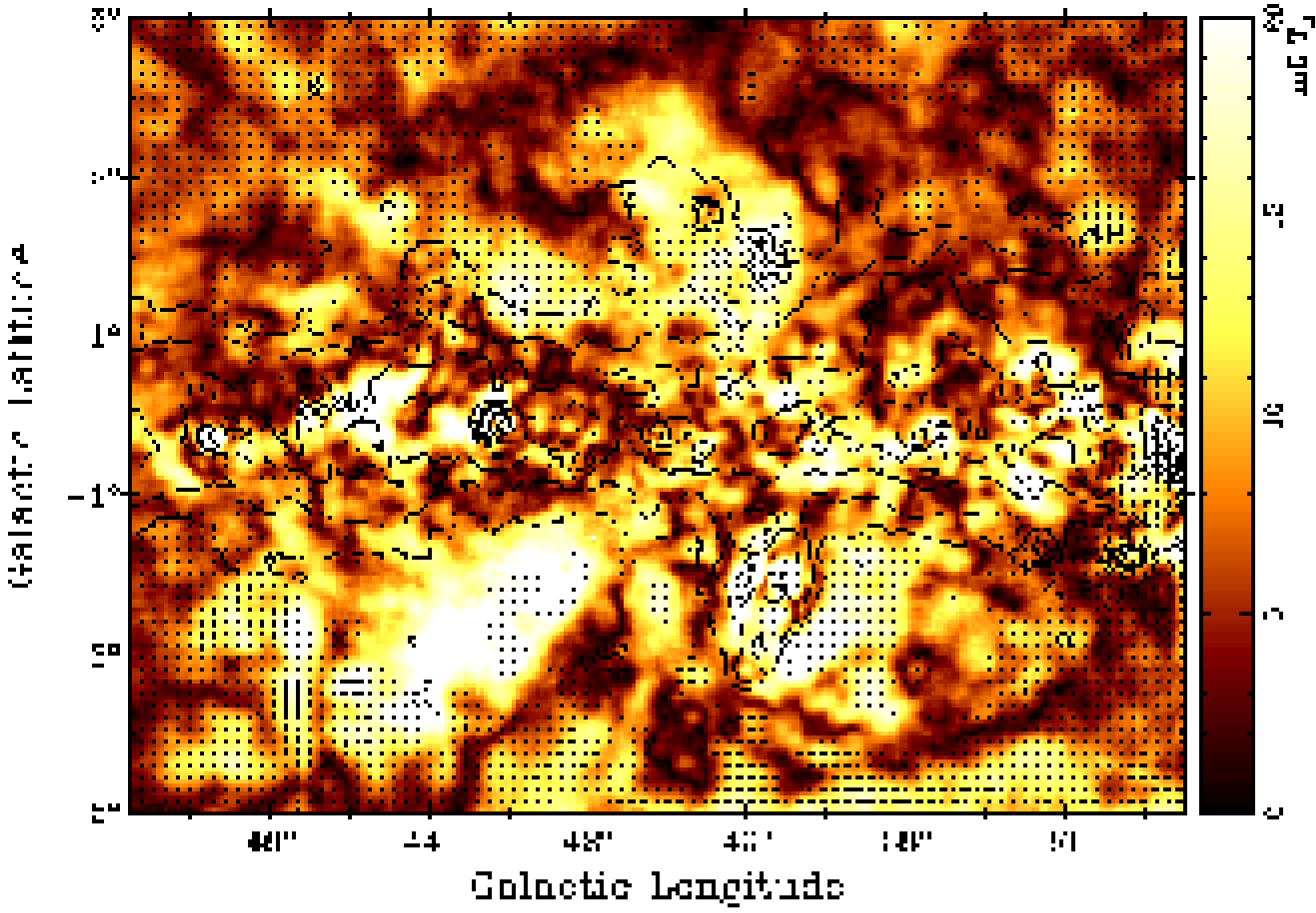}}
\caption{Zero-level restored $PI$ maps overlaid by total intensity contours and 
bars showing B-vectors. The contours run in steps of 
$2^n\times50$~mK~$T_{\rm B}$ 
with $n=0,\,1,\,2,\,\ldots$. The lengths of B-vectors are proportional to $PI$. 
Below an intensity cutoff of 2.5~mK~$T_{\rm B}$ (5$\times$rms-noise) no bars 
are shown.}
\label{sp12}
\end{figure*}

\begin{figure*}[!htbp]
\centering
\resizebox{0.95\textwidth}{!}{\includegraphics[angle=-90]{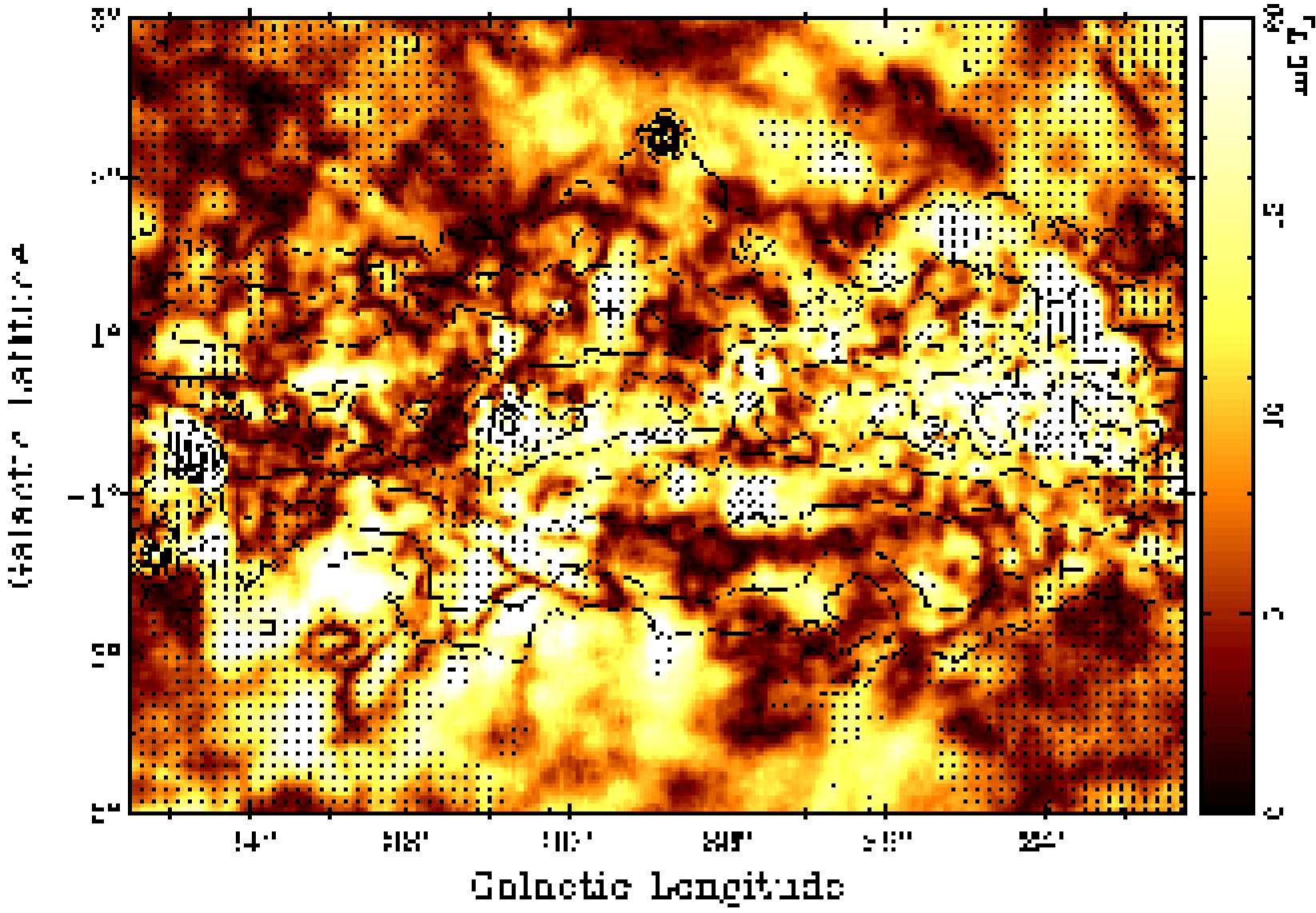}}
\resizebox{0.95\textwidth}{!}{\includegraphics[angle=-90]{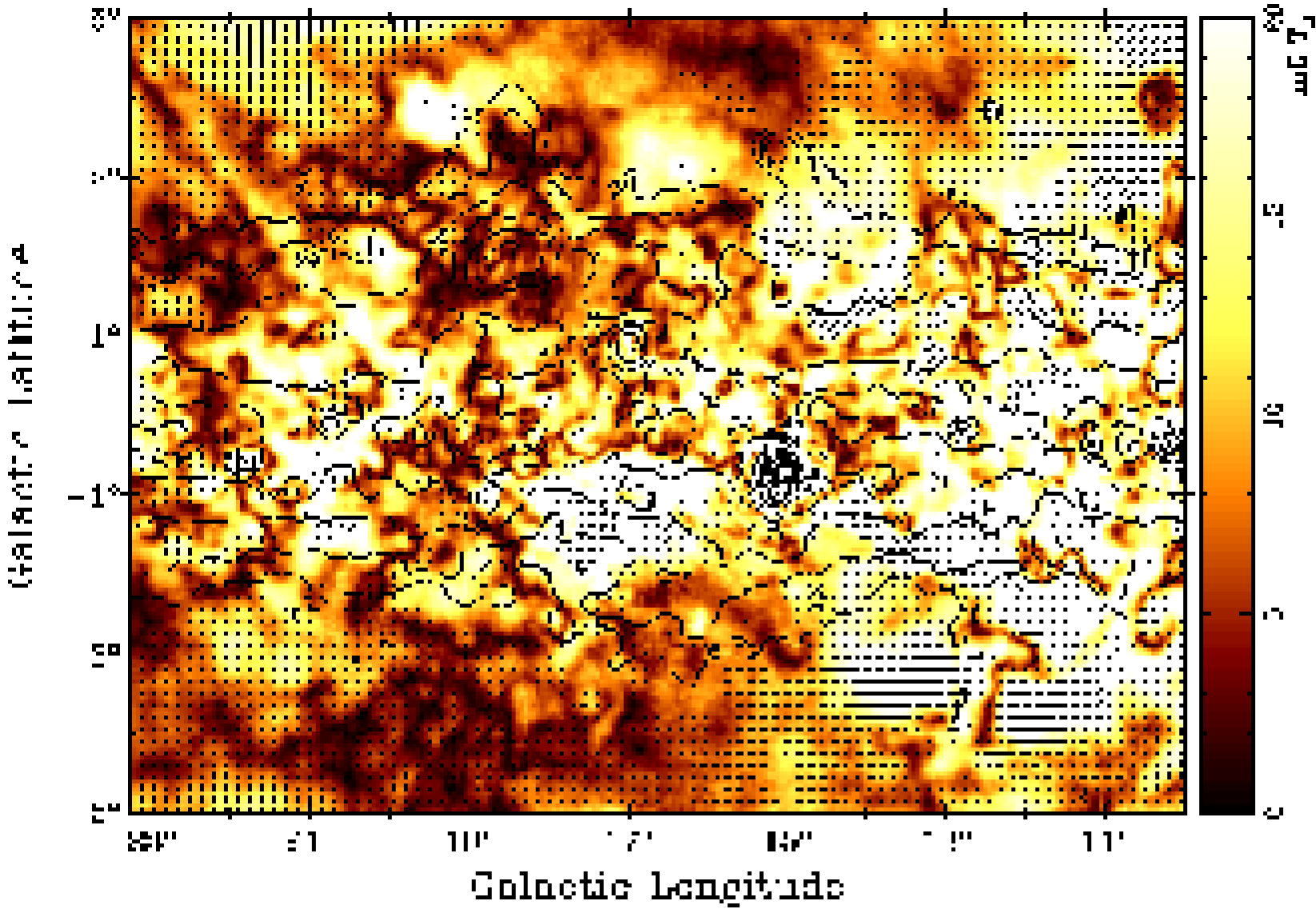}}
\caption{The same as Fig.~\ref{sp12} for the lower longitude regions.}
\label{sp34}
\end{figure*}

\begin{figure}[!htbp]
\centering
\includegraphics[width=5cm,angle=-90]{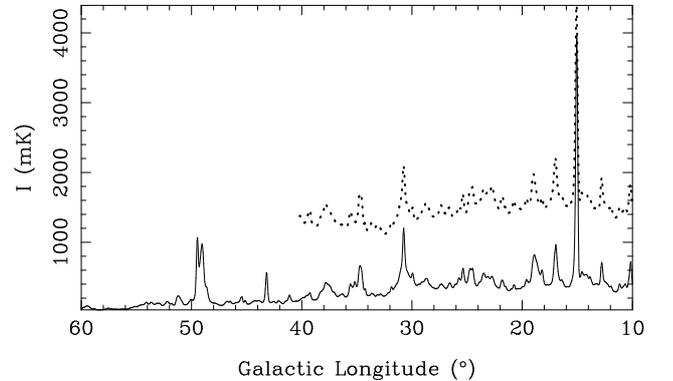}
\caption{Total intensity $I$ along Galactic longitude averaged within 
$b=\pm2\degr$. The solid line shows the $\lambda$6~cm survey data and the 
dotted line the Parkes survey \citep{hcs78} data with an offset of about 
1000~mK.}
\label{prof66}
\end{figure}

$I$, $U$ and $Q$ maps are shown in Fig.~\ref{g35all}. $PI$ maps calculated 
from the zero-level restored $U$ and $Q$ maps are shown in Figs.~\ref{sp12} and 
\ref{sp34}. The $I$ maps show strong diffuse emission concentrated along the 
ridge of the Galactic plane, where a high density of discrete sources such as 
\ion{H}{II} regions and supernova remnants (SNRs) is seen. The emission 
increases towards lower longitudes. The longitude profile for $I$ averaged 
within $b=\pm2\degr$ is compared with that of the Parkes $\lambda$6~cm survey 
\citep{hcs78} in Fig.~\ref{prof66}. The total intensity structures agree quite 
well. The TT-plot yields an intensity ratio of $1.0\pm0.1$ and an arbitrary 
offset of about 1000~mK (which was added to the Parkes data release). This 
proves in addition that both $\lambda$6~cm surveys are consistent. 

Polarized emission shows different structures compared to those seen in total 
intensity (Figs.~\ref{sp12} and \ref{sp34}). There is strong polarized emission 
even far above the plane, which is similar to that seen in the Effelsberg 
$\lambda$11~cm survey maps \citep{jfr87,drrf99}. However, the polarized 
structures revealed by these two surveys do not correspond to each other, 
because of different base-line settings and polarization horizons. A 
$10\degr$-region centred at $l=50\degr$, $b=0\degr$ has been observed as part 
of the Toru{\'n} $\lambda$6~cm survey project \citep{rck+06}. The most 
prominent polarized structures, such as the patch at $l=49\fdg80$, $b=1\fdg15$, 
agree with our $\lambda$6~cm polarization survey. 

It is evident that the B-vectors do not align well with the Galactic plane, 
which is either intrinsic or caused by strong Faraday rotation along the 
line-of-sight. 

The most prominent structures are,
\begin{enumerate}
\item Structures visible in both total and polarized intensities: These are 
      point sources and SNRs. Particular bright are G34.7$-$0.4 (W~44), 
      G39.7$-$2.0 (W~50) and G54.4$-$0.3 (HC~40). A weak polarized filament is 
      traced from $l=23\fdg5$, $b=5\degr$ towards lower latitudes. This 
      filament was previously identified as an extension of the North Polar 
      Spur (NPS) by total intensity observations at $\lambda$21~cm by 
      \citet{sr79}. For the first time we are able to trace this filament also 
      in polarization.

\item Polarized patches without corresponding total intensities: The patches 
      span wide scales from tens of arcmin to several degrees. Most of them are 
      seen outside of the Galactic plane ridge. More patches are seen at low 
      longitudes than at large longitudes. These patches are probably produced 
      by turbulent fields as discussed below.

\item Narrow depolarization regions commonly called ``canals": They are most 
      pronounced in the region around $11\degr<l<15\degr$. These canals are 
      not related to total intensity structures. As demonstrated later an 
      abrupt polarization angle change by about $0.6\pi$ over 
      $3\arcmin$ can cause such canals. 

\item Large depolarized regions we call ``voids": Most voids do not correspond 
      to a total intensity minimum or other feature. An example is seen at 
      $l=40\fdg9$, $b=-4\fdg2$. These regions were modeled in Sect.~5.2 by the 
      Faraday Screen model already used in Papers~I and II.
\end{enumerate}

\section{Discrete objects}

For point-like or compact sources, flux densities were obtained by fitting a 
two-dimensional elliptical Gaussian. A list of compact sources from the entire 
survey will be presented in a forthcoming paper. Studies of SNRs located in 
this area will also be presented in subsequent papers. Other discrete objects 
such as the NPS extension as well as \ion{H}{II} regions will be discussed in 
this section. Polarized structures with no counterparts in total intensity are 
analysed in Sect.~5.

\subsection{Extended sources in general}

Prominent extended sources are either SNRs or \ion{H}{II} regions. Their 
spectra differ and can be used to identify them. Shell-type SNRs are 
non-thermal sources with spectral indices close to $\alpha \sim -0.5$, whereas 
\ion{H}{II} regions are thermal with nearly flat spectra. The spectra of 
strong sources can be determined by their integrated flux density at several 
frequencies. For weak sources, whose flux density cannot be determined very 
well, we investigated their spectra by TT-plots and fit their slopes. The 
maps throughout the paper were smoothed to $9\farcm5$ required to perform 
TT-plots. The TT-plot method largely circumvents the influence of background 
emission differences amongst maps from different surveys. Plerions and pulsar 
wind nebulae (PWNe) are non-thermal SNRs, but exhibit flat spectra similar to 
\ion{H}{II} regions. However, all types of SNRs are polarized and show 
X-ray emission, while \ion{H}{II} regions are usually associated with 
strong infrared sources. The ratio of IRAS 60~$\mu$m to radio continuum 
intensity is usually much larger for \ion{H}{II} regions than for SNRs 
\citep[e.g. ][]{frs87}. We used X-ray images from the ROSAT all-sky 
survey\footnote{http://www.xray.mpe.mpg.de/cgi-bin/rosat/rosat-survey} 
and high-resolution IRAS 60~$\mu$m maps \citep{ctpb97} to distinguish between 
SNRs and \ion{H}{II} regions.

\subsection{\ion{H}{II} regions}

\ion{H}{II} regions reside predominantly in spiral arms and thus numerous 
\ion{H}{II} regions are seen in the present survey section as the line-of-sight 
intersects many spiral arms \citep{hhs09}. Most of the known \ion{H}{II} 
regions have been catalogued by \citet{pbd+03} and \citet{hhs09}. They are 
visible in the $I$ maps. Some emission complexes, as for example the W~51 
complex centred at $l=49\fdg2$, $b=-0\fdg4$, host many individual \ion{H}{II} 
regions surrounded by diffuse emission. Such a study is beyond the scope of 
this paper. The $\lambda$6~cm survey reveals some previously unknown large and 
weak \ion{H}{II} regions as demonstrated by \citet{ssh+08}. However, some 
efforts are required to discover new \ion{H}{II} regions in the current region 
because of obscuration by strong diffuse emission. Therefore we limit our 
investigation to two objects out of the Galactic plane or longitudes close to 
$l=60\degr$, where the diffuse emission is weaker.  
  
\begin{figure}[!htbp]
\centering
\includegraphics[width=8cm,angle=-90]{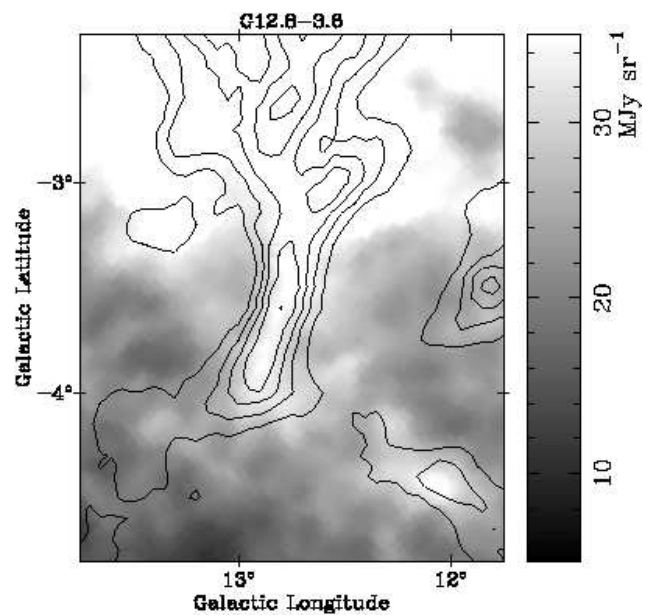}
\caption{IRAS 60~$\mu$m greyscale image showing the \ion{H}{II} region 
G12.8$-$3.6. Contours display total intensities at $\lambda$6~cm 
starting at 10~mK~$T_{\rm B}$ and running in steps of 10~mK~$T_{\rm B}$.}
\label{hii12.8r}
\end{figure}

\begin{figure}[!htbp]
\centering
\includegraphics[width=6.5cm,angle=-90]{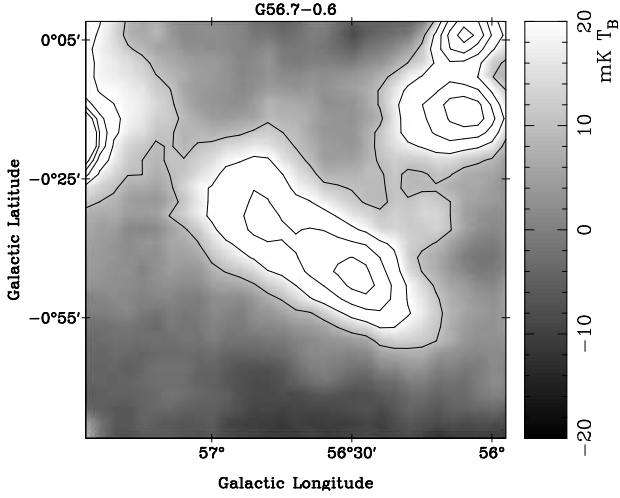}
\caption{$\lambda$6~cm total intensity image for the newly identified 
\ion{H}{II} region G56.7$-$0.6 shown in greyscale and contours, 
which start at 10~mK~$T_{\rm B}$ running further in steps of 
10~mK~$T_{\rm B}$. }
\label{hii56.7}
\end{figure}

We identified two thermal structures in this survey region, which could be new 
\ion{H}{II} regions. The first one, G12.8$-$3.6, at $l=12\fdg75$, $b=-3\fdg55$, 
has a size of about $100\arcmin\times24\arcmin$ (Fig.~\ref{hii12.8r}). Its 
integrated flux density is difficult to measure because of its complex 
environment. The spectral index from a TT-plot between the $\lambda$6~cm and 
Effelsberg $\lambda$11~cm data is $\alpha=0.15\pm0.29$, which indicates that 
the source is thermal. The object is at the latitude limit of the Effelsberg 
$\lambda$21~cm survey, which was not used for the analysis. The thermal nature 
of G12.8$-$3.6 is further corroborated by infrared emission at 60~$\mu$m 
coinciding exactly with the radio emission (Fig.~\ref{hii12.8r}). The second 
source, G56.7$-$0.6, located at $l=56\fdg70$, $b=-0\fdg60$, has an apparent 
size of about $65\arcmin\times27\arcmin$ (Fig.~\ref{hii56.7}). The flux 
densities measured from the $\lambda$6~cm survey and Effelsberg $\lambda$11~cm 
and $\lambda$21~cm surveys are 1.76$\pm$0.09~Jy, 1.32$\pm$0.07~Jy, and 
1.96$\pm$0.11~Jy, respectively, which yields a spectral index of 
$\alpha=0.08\pm0.06$. Unfortunately we cannot find exciting stars for both 
structures from other database. 

Another object G57.7+0.3 was already proposed by \citet{ss84} to be a 
\ion{H}{II} region, but was not catalogued by \citet{pbd+03}. We measured flux 
densities at $\lambda$6~cm and Effelsberg $\lambda$11~cm and $\lambda$21~cm 
surveys to be 2.51$\pm$0.13~Jy, 2.33$\pm$0.13~Jy, and 2.54$\pm$0.20~Jy, 
respectively. The spectral index is $\alpha=0.00\pm0.07$. Strong 
infrared emission is associated, we therefore confirm this source to be a 
\ion{H}{II} region. 

\subsection{The southern North Polar Spur extension}

\begin{figure}[!htbp]
\centering
\includegraphics[width=7cm,angle=-90]{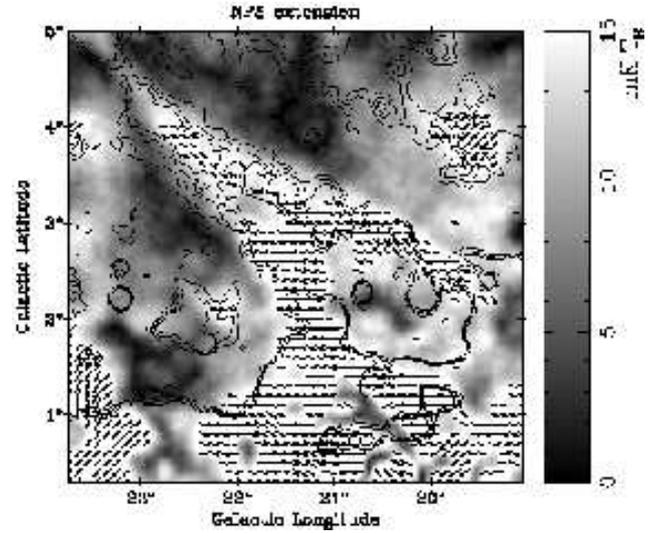}
\caption{Polarization image for the southern NPS extension. Contours show 
total intensities and bars represent B-vectors. Contours start at 
5~mK~$T_{\rm B}$ and run in steps of 4.5~mK~$T_{\rm B}$.}
\label{nps}
\end{figure}

The North Polar Spur (NPS) is most probably a local and very old SNR 
\citep{sal83}. Its filamentary shell structure should not be affected when 
running across the Galactic plane. However, the strong emission from unrelated 
more distant Galactic plane emission causes confusion that it becomes 
very difficult to trace the NPS in this area. This was already demonstrated by 
$\lambda$21~cm observations by \citet{sr79}. At $\lambda$6~cm confusion is 
less strong, so that the filamentary southern NPS extension can be traced from 
$l=23\degr$, $b=5\degr$ towards a latitude of about $2\fdg5$ in total 
intensity. At lower latitudes it is confused with strong diffuse emission 
(Fig.~\ref{nps}). The NPS extension can be traced also by its polarization 
towards a lower latitude of about $0\fdg5$ and eventually to a latitude of 
about $-3\degr$. The average polarized intensity is about 15 mK~$T_{\rm B}$. 
At about $l=22\fdg4$, $b=3\fdg6$ the NPS splits into two filaments as seen at 
$\lambda$21~cm in total intensity \citep{sr79}. This is also seen at 
$\lambda$6~cm, where polarization from the two filaments is seen as well. The 
southern filament is longer and the northern part shorter at $\lambda$6~cm in 
total intensity than at $\lambda$21~cm. The southern filament is inclined by 
about $50\degr$ relative to the Galactic plane, while the northern filament is 
inclined by $30\degr$. The B-vectors are roughly orientated tangential to the 
northern filament. For the southern filament the B-vector direction is almost 
parallel to the Galactic plane for latitudes below about $2\fdg5$.

\citet{wol07} has modeled the NPS by two local shells mainly based on DRAO 
1.4~GHz all-sky polarization survey data \citep{wlrw06}. However, this 
polarization survey suffers strong depolarization for latitudes below about 
$30\degr$, which prevents tracking the NPS towards low latitudes. This global 
model does not consider splitting of the filamentary shell as observed. 
Follow-up observations extending the $\lambda$6~cm survey for the higher 
latitude NPS regions are on the way, that we postpone any further discussion.

\section{Polarized structures}

\subsection{General considerations}

The maximum theoretical polarization percentage of synchrotron emission is 
about 75\%. However, random magnetic fields may reduce the polarization 
percentage significantly. For a ratio of random to regular magnetic fields of  
1.5, the intrinsic percentage polarization reduces to about 30\% \citep{srwe08}.
We will use this value throughout this paper. Two mechanisms may further smear 
out polarized emission: depth polarization and beam depolarization.   

Depth depolarization occurs when polarized synchrotron emission is mixed with 
thermal gas. Polarized emission originating at different distances along the 
line-of-sight experiences a different amount of Faraday rotation and thus 
different polarization angles. Adding polarized emission components along the 
line-of-sight causes a reduction up to an entire cancellation of polarization. 
Quantitatively the amount of depolarization $p$, i.e. the ratio of the observed 
to intrinsic polarized percentage, can be written as \citep{sbs+98}, 
\begin{equation}\label{ddp}
p=\frac{1-\exp(-S)}{S},
\end{equation} 
where 
$S=2\sigma_{\rm RM}^2\lambda^4-2{\mathbf i}\mathcal{R}\lambda^2$, 
and $\sigma_{\rm RM}$ is the RM scattering along the line-of-sight.
 If $\sigma_{\rm RM}=0$, Eq.~(\ref{ddp}) simplifies to 
$p=\sin(\mathcal{R}\lambda^2)/\mathcal{R}\lambda^2$, which is frequently used. 

Beam depolarization occurs by Faraday rotation in front of the 
synchrotron-emitting medium. RM variations on scales smaller than the beam 
width result in polarization angle difference. By integration across the beam 
the observed polarization is reduced. Following \citet{sbs+98} the 
depolarization factor $p$ can be expressed as,
\begin{equation}\label{bdp}
p=\exp(-2\sigma_{\rm RM,\,\bot}^2\lambda^4),
\end{equation}  
where $\sigma_{\rm RM,\,\bot}$ is the RM variance across the beam. A rough 
estimate for $\sigma_{\rm RM,\,\bot}$ can be given as \citep{gdm+01}, 
\begin{equation}\label{sigrm}
\sigma_{\rm RM,\,\bot}=\frac{Kn_e b L}{2\sqrt{3}}
                       \left(\frac{l}{L}\right)^{1/2}, 
\end{equation} 
where $K$ is a constant, $n_e$ is the electron density, $b$ is the strength of 
the random magnetic field, $L$ is the depth through the Faraday rotating 
medium, and $l$ is the coherent length of the random magnetic fields. The 
quantity $Kn_ebL$ is roughly equal to $\sigma_{\rm RM}$ and $l$ is about the 
spatial scale resolved by the beam for a distance $L$.

\begin{figure}[!htbp]
\centering
\resizebox{0.46\textwidth}{!}{\includegraphics[angle=-90]{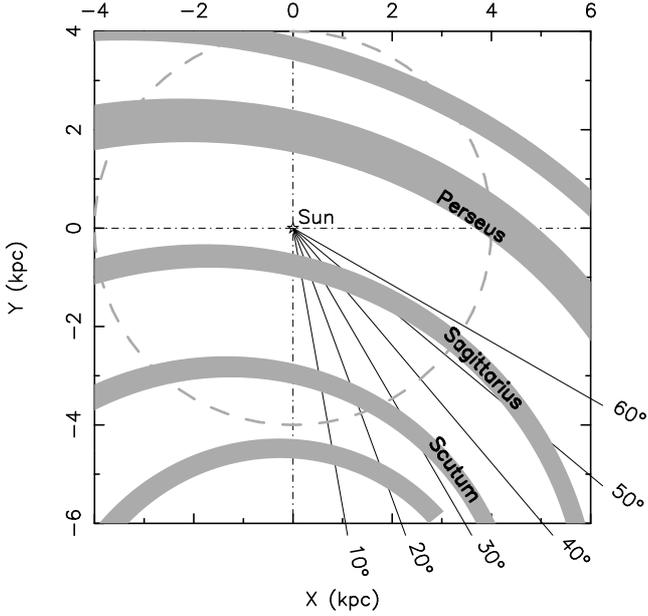}}
\caption{Sketch of the location of Galactic spiral arms taken from the 
polynomial spiral arm model of \citet{hhs09}. The longitude range of the 
present survey section is indicated. The dashed circle has a radius of 4~kpc, 
and indicates the polarization horizon at $\lambda$6~cm in the Galactic plane.}
\label{arm}
\end{figure}

For this $\lambda$6~cm survey section the polarization horizon in the Galactic 
plane was estimated to be about 4~kpc based on simulations as described by 
\citet{sr09} (see Fig.~\ref{dph}). At distances larger than 4~kpc the average 
pulsar RM is about 200~rad~m$^{-2}$ and the RM variance along the line-of-sight 
is about 400~rad~m$^{-2}$ (Fig.~\ref{psrrm}). With these values a 
depolarization factor $p$ of about 0.2 is calculated using Eq.~(\ref{ddp}), 
which means a quite significant depolarization. The extended \ion{H}{II} region 
W~35 at $l=18\fdg45$, $b=2\fdg05$ is located at a distance of about 
2.9~kpc \citep{mrr87} and does almost not modulate polarization, which is 
consistent with the simulated polarization horizon. Synchrotron emission 
originates at most in spiral arms, so that we see polarized emission from the 
Sagittarius and the Scutum arms for the inner region and only from the 
Sagittarius arm for the outer region (see Fig.~\ref{arm}). For the longitude 
range of $50\degr<l<60\degr$ the line-of-sight is almost tangential to the 
Sagittarius arm (Fig.~\ref{arm}). It is commonly assumed that the large-scale 
magnetic fields follow the spiral arms \citep{hml+06} and thus RM increases. 
This results in a higher degree of depolarization in this direction.

The $\lambda$6~cm maps of this survey section show numerous depolarization 
features such as voids and canals (Figs.~\ref{sp12} and \ref{sp34}). Within 
the polarization horizon, depth depolarization is not important, because 
absolute pulsar RMs and their fluctuations are small. If beam depolarization is 
significant, we calculate that for a depolarization factor of $p = 0.1$ a RM 
variance across the beam of about 275~rad~m$^{-2}$ is required 
(see Eq.~(\ref{bdp})). According to Eq.~(\ref{sigrm}) 
$\sigma_{\rm RM,\,\bot} \approx \sigma_{\rm RM}\sqrt{l/L}/2\sqrt{3}$. Pulsar 
RMs give $\sigma_{\rm RM}\leq400$~rad~m$^{-2}$, and $l/L$ is about the beam 
size. This implies $\sigma_{\rm RM,\,\bot}\approx6$~rad~m$^{-2}$, or $p = 1$. 
We conclude that beam depolarization can be largely neglected.  

Most of the Faraday Screens discussed in Paper~I and II cause no large 
depolarization 
of background emission, but rotate the polarization background position angle. 
After rotation the polarized background and foreground emission have different 
polarization angles compared to their surroundings. The observed sum of both 
components then shows a decrease in polarized intensity. A simple model was 
described and extensively used in Papers~I and II. The positions were defined 
as ``on" if the line-of-sight passes the Faraday Screen and ``off" otherwise. 
In general, the polarized intensity and angle towards ``on" and ``off" 
positions can be expressed as, 
\begin{equation}\label{fsmodel}
\left\{
\begin{array}{rcl}
PI_{\rm on}&=&\sqrt{PI^2_{\rm bg}+PI^2_{\rm fg}+2PI_{\rm bg}PI_{\rm fg}
               \cos(2\psi_{\rm bg}+2\psi_{\rm s}-2\psi_{\rm fg})}\\[4mm]
\psi_{\rm on}&=&\displaystyle{\frac{1}{2}{\rm atan}
      \frac{PI_{\rm bg}\sin(2\psi_{\rm bg}+2\psi_{\rm s})+
            PI_{\rm fg}\sin2\psi_{\rm fg}}
           {PI_{\rm bg}\cos(2\psi_{\rm bg}+2\psi_{\rm s})+
            PI_{\rm fg}\cos2\psi_{\rm fg}}}\\[5mm]
PI_{\rm off}&=&\sqrt{PI^2_{\rm bg}+PI^2_{\rm fg}+2PI_{\rm bg}PI_{\rm fg}
               \cos(2\psi_{\rm bg}-2\psi_{\rm fg})}\\[4mm]
\psi_{\rm off}&=&\displaystyle{\frac{1}{2}{\rm atan}
      \frac{PI_{\rm bg}\sin2\psi_{\rm bg}+PI_{\rm fg}\sin2\psi_{\rm fg}}
           {PI_{\rm bg}\cos2\psi_{\rm bg}+PI_{\rm fg}\cos2\psi_{\rm fg}}}
\end{array}
\right.
\end{equation} 
A Faraday Screen causes an angle rotation $\psi_{\rm s}={\rm RM_s}\lambda^2$. 
The subscripts ``fg" and ``bg" denote the foreground and background components, 
respectively. If only data at one frequency are available, the model may be 
simplified  that $\psi_{\rm bg}=\psi_{\rm fg}=\psi_{\rm off}$. For the 
anti-centre regions as discussed in Paper~I and II, the dominating emission 
originates from the disk field, where the magnetic field is in general parallel 
to the Galactic plane. This means $\psi_{\rm off}$ is close to $0\degr$. 

\subsection{Polarized patches}

Regions showing strong polarized emission are predominantly ``patches", such 
as the extended feature centred at $l=43\fdg80$, $b=-2\fdg90$. Most of these 
patches do not have corresponding total intensity emission. The patches show 
either intrinsic polarization from a region in interstellar space or they are
caused by Faraday Screen modulation of polarized background emission. 

We consider a Faraday Screen origin for the patches in general as unlikely. In 
this scenario polarization angles of the foreground and background components 
must differ. Faraday Screens have to rotate the polarized background, so that a 
reduction of the angle difference to the foreground component is obtained. In 
this case the polarized emission appears to be stronger compared to its 
surroundings. The survey maps show that the polarized intensity of the patches 
is in most cases much larger compared to their surroundings. In general the 
ratio of $PI_{\rm off}/PI_{\rm on}$ is estimated to be less than about 0.3. 
If foreground and background polarized intensities are nearly identical, their 
angles should satisfy 
$\psi_{\rm s}\approx\psi_{\rm fg}-\psi_{\rm bg}\approx90\degr$ 
(see Eq.~(\ref{fsmodel})) towards all the patches. It is unlikely that this is 
true over a wide area. If foreground and background intensities differ 
significantly a ratio of $PI_{\rm off}/PI_{\rm on}$ less than 0.3 can not be 
caused by Faraday Screen action at all. The RMs of the Faraday Screens have to 
be less than 200~rad~m$^{-2}$ as inferred from pulsar RMs (Fig.~\ref{psrrm}). 
This results in a maximum angle rotation by a Faraday Screen of about $45\degr$ 
at $\lambda$6~cm. Moreover it is generally believed that the magnetic field 
perpendicular to the plane is much smaller than that parallel to the plane 
\citep[e.g. ][]{hq94,hmq99,srwe08}. Thus angle differences between background 
and foreground shall approach zero. Therefore the patches are unlikely to be 
produced by Faraday Screens.

Thus we consider the patches in their majority as intrinsically polarized 
features. Since they have no correspondence in total intensity, they are not 
caused by polarized objects such as SNRs. Instead we consider the patches as 
diffuse synchrotron emission components originating from turbulent field cells, 
whose power spectrum follows a power-law with a non-zero spectral index. The 
synchrotron emission observed from various turbulent field cells with different
magnetic field orientations is smooth or almost structureless in total 
intensity. Polarized emission, however, shows irregularities or inhomogeneities 
depending on the field orientation in the different cells and their correlation.
 As shown by high angular resolution simulations by \citet{sr09}, turbulent 
magnetic fields with a Kolmogorov-like spectrum produce polarized features 
without corresponding total intensity. In the anti-centre region the cosmic ray 
electron density is low and the polarized intensities contributed by individual 
turbulent field cells is small as well. Therefore we do not expect to see many 
polarized patches, which is consistent with the $\lambda$6~cm observations 
presented in Papers~I and II. At K-band the polarization horizon is large 
everywhere in the Galaxy. Many turbulent cells along the line-of-sight average 
out all distinct structures and thus individual polarized patches are not 
expected to be seen in the WMAP maps \citep{hwh+09}, what is in agreement with 
the observations.

\begin{figure}[!htbp]
\centering
\includegraphics[width=5.5cm,angle=-90]{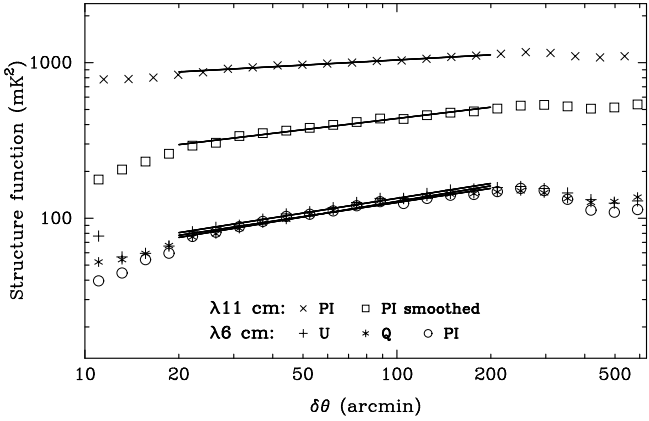}
\caption{$U$, $Q$ and $PI$ structure functions for the present $\lambda$6~cm 
survey section. The structure functions calculated for smoothed ($9\farcm5$) and original ($4\farcm3$)
angular resolution $\lambda$11~cm $PI$ data are also shown.} 
\label{sf}
\end{figure}

We apply structure functions to study the properties of the polarized patches 
following the approach by \citet{sr09}. It is yet unknown how the turbulent 
magnetic field properties are reflected in structure functions. The structure 
functions are not influenced by missing large-scale components, so that the 
analysis was based on the original maps. The structure functions for $U$, $Q$ 
and $PI$ at $\lambda$6~cm are shown in Fig.~\ref{sf}. For comparison we 
retrieved the Effelsberg $\lambda$11~cm $U$ and $Q$ data for this survey 
region, smoothed them to the same angular resolution as the $\lambda$6~cm survey 
and re-calculated polarized intensity. The structure function for the smoothed 
$PI$ at $\lambda$11~cm as well as the original with an angular resolution of 
$4\farcm3$ are also shown in Fig.~\ref{sf}. All structure functions can be 
described by a power-law. The spectral index is about 0.3 for the $9\farcm5$
angular resolution data and 0.1 for the $4\farcm3$ resolution $\lambda$11~cm 
$PI$ data. The spectral index refers to spatial scales between about 
$20\arcmin$ and $3\degr$, which corresponds to sizes of about 20~pc and 200~pc 
for the polarized patches in case their distance is 4~kpc. 

\begin{figure}[!htbp]
\centering
\resizebox{0.4\textwidth}{!}{\includegraphics[angle=-90]{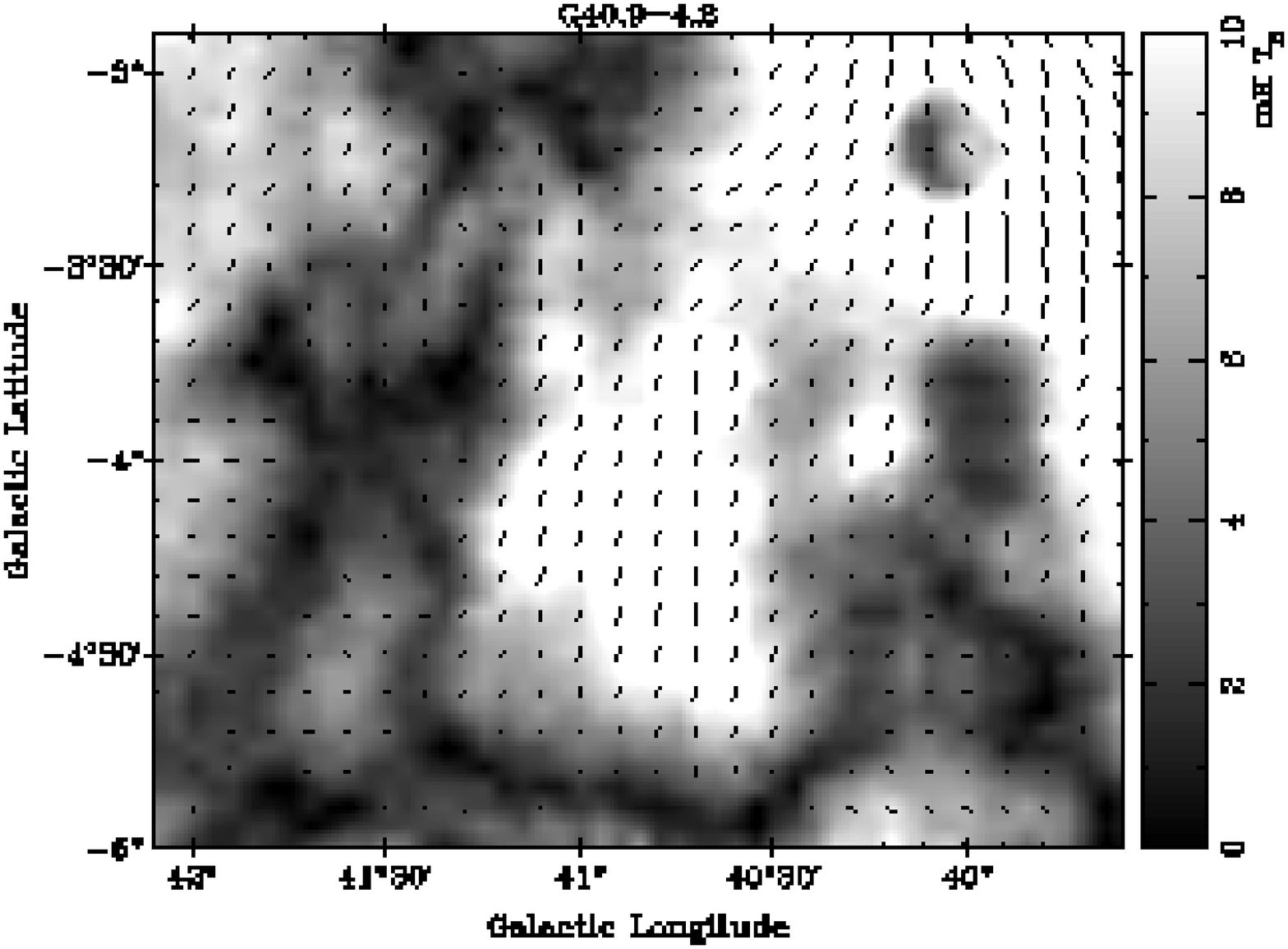}}
\resizebox{0.4\textwidth}{!}{\includegraphics[angle=-90]{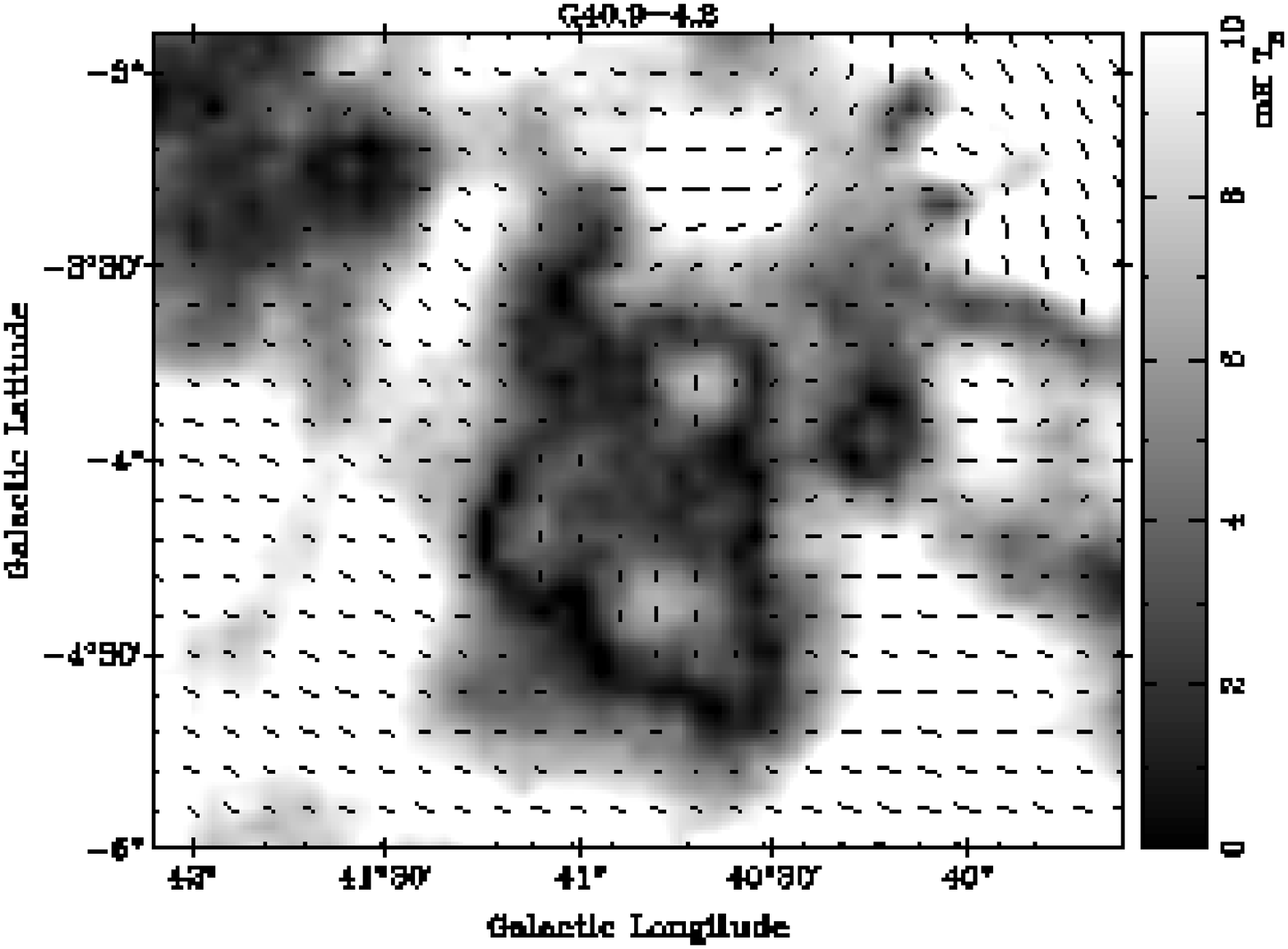}}
\resizebox{0.4\textwidth}{!}{\includegraphics[angle=-90]{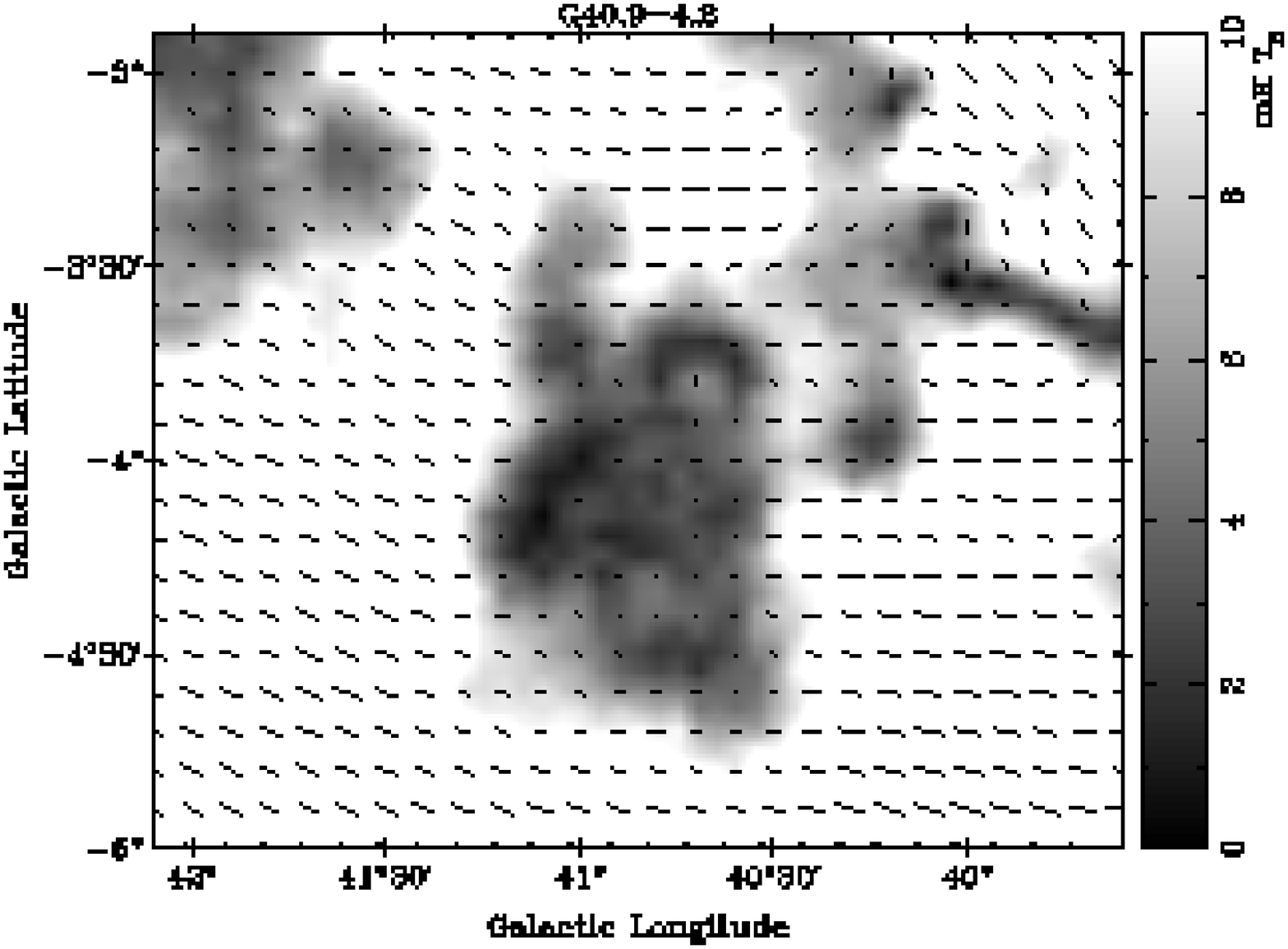}}
\caption{Polarized intensity maps for the void G40.9$-$4.2. The {\it 
upper panel} shows the original map, the {\it middle panel} shows the map with 
the zero-level restored with a spectral index of $\beta_{PI}=-2.7$ according to 
the restoration model discussed in Sect.~\ref{spmodel}. The {\it lower panel} 
shows the map with the zero-level restored using the lowest possible spectral 
index of $\beta_{PI}=-3.1$ (see Fig.~\ref{beta_kka}). $PI$ is always shown in 
grey scale, and bars indicate polarization B-vectors.}
\label{voids_40.9}
\end{figure}

\begin{figure}[!htbp]
\centering
\resizebox{0.4\textwidth}{!}{\includegraphics[angle=0]{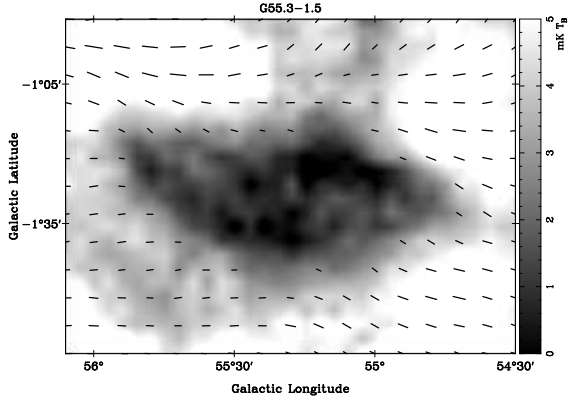}}
\caption{Polarization void G55.3$-$1.5. $PI$ is shown in grey scale, and bars 
indicate polarization B-vectors.}
\label{voids_55.3}
\end{figure}

The morphology of patches resulting from turbulent magnetic field cells should 
vary with frequency and resolution. The statistical properties such as the 
slope of the structure function should be independent on frequency. However, 
statistics depend on angular resolution. For low angular resolutions polarized 
patches from nearby turbulent field cells dominate, which will cause a 
steepening of the spectrum of the structure function. The slope of the 
$\lambda$11~cm structure functions in logarithmic scale changed from about 0.1 
to 0.24 after smoothing, consistent with the above expectation. The slope of 
0.24 for smoothed $\lambda$11~cm $PI$ is almost the same, compared to 0.3 
as calculated for the $\lambda$6~cm polarization data. This result supports our 
argument that polarized patches reflect turbulent field cells. The slight 
discrepancy between $\lambda$11~cm and $\lambda$6~cm may stem from differences 
of their polarization horizon and the corresponding different emission volumes. 
Simulations were made as described in Sect.~\ref{dph_sim}, which indicate 
that the polarization horizon at $\lambda$11~cm is about 1--3~kpc compared to 
about 4~kpc at $\lambda$6~cm.

\subsection{Large polarization voids}

We define ``voids" as large almost completely depolarized regions without 
correspondence in total intensity. Previously \citet{st07} reported on
regions of about $10\degr$ in size, where the density of extragalactic sources
from the NVSS is reduced by a factor 2--4. The size of these polarization 
shadows indicates a local origin for the depolarization and \citet{st07} 
discussed several possibilities. The voids seen at $\lambda$6~cm could 
neither be caused by depth nor by beam depolarization as discussed above. 
Instead we use the Faraday Screen model described above to understand the 
polarization voids. 

The void G40.9$-$4.2, located at $l=40\fdg85$, $b=-4\fdg15$, has a size of 
about $55\arcmin\times40\arcmin$ (Fig.~\ref{voids_40.9}). It is almost entirely 
depolarized towards its centre. To demonstrate the importance of absolute 
calibration when discussing ``voids'' we show the original polarized intensity 
map and the map with large-scale emission added in Fig.~\ref{voids_40.9}. The 
void G40.9$-$4.2 appears as a polarized emission feature in the original map 
and turns into a depression when large-scale components were added. Instead of 
using the spectral index of $\beta_{PI}=-2.7$ we also show the large-scale 
level restored with a spectral index of $\beta_{PI}=-3.1$ (lower panel in 
Fig.~\ref{voids_40.9}), which is the lower spectral index limit for this 
longitude range (see Sect.~~\ref{spmodel}). As shown in Fig.~\ref{voids_40.9} 
the ``void'' slightly shrinks in size, but its morphology remains almost 
unchanged. This demonstrates that the uncertainties related to the spectral 
index determination do not have a large influence on the analysis of 
polarized structures.

Another outstanding example of a void is G55.3$-$1.5 located at $l=55\fdg30$, 
$b=-1\fdg50$ (Fig.~\ref{voids_55.3}). Its size is about 
$51\arcmin\times22\arcmin$. The polarized emission drops close to zero towards 
its centre direction. In this case the large-scale corrections of $U$ and $Q$ 
are very close to zero and therefore the maps with and without restoration are 
almost identical.

According to the Faraday Screen model (Eq.~(\ref{fsmodel})) the foreground and 
background polarized intensity should be almost equal and the polarization 
angle rotation by the Faraday Screen is 
$\psi_{\rm s}=90\degr+\psi_{\rm fg}-\psi_{\rm bg}$. The ``off" polarization 
angle should be $\psi_{\rm off}=(\psi_{\rm fg}+\psi_{\rm bg})/2$. For both 
voids we measured $\psi_{\rm off}\approx160\degr$, indicating 
$\psi_{\rm fg}+\psi_{\rm bg}=320\degr$. As both $\psi_{\rm fg}$ and 
$\psi_{\rm bg}$ are in the range between $0\degr$ and $180\degr$, we obtain 
$-40\degr\leq \psi_{\rm fg}-\psi_{\rm bg}\leq40\degr$, and subsequently 
$50\degr\leq\psi_{\rm s}\leq130\degr$. The RM of the Faraday Screens needs to 
be about $220$~rad~m$^{-2}$ to account for $\psi_{\rm s}=50\degr$, which is 
about the RM maximum constrained by pulsar RM in this area (Fig.~\ref{psrrm}). 
Larger values could not be entirely ruled out, because there are no pulsar RMs 
measured in the direction of the voids. In case of $\psi_{\rm s}=50\degr$ the 
angle difference between the foreground and background emission is about 
$40\degr$. For a smaller angle difference the RM of the Faraday Screens must be 
larger. 

In case the polarized emissivity is uniform along the line-of-sight, the 
distance to the Faraday Screens is about half of the polarization horizon, or 
about 2~kpc (Fig.~\ref{dph}). This yields a mean size of about 28~pc for 
G40.9$-$4.2, and about 21~pc for G55.3$-$1.5. The $\lambda$6~cm brightness 
temperature contributed by the warm thermal gas from the Faraday Screen can be 
written as $T=T_e\tau=0.13n_e^2L$~mK. Here the opacity $\tau$ was calculated 
according to the formula by \citet{rwbook}, where $n_e$ is the electron 
density in cm$^{-3}$ and $L$ is the size of the Faraday Screen in pc. The 
electron temperature $T_e$ was taken to be 8000~K. The non-detection of the 
Faraday Screen in total intensity provides an upper limit for the electron 
density of about 1.2~cm$^{-3}$ for G40.9$-$4.2, and 1.4~cm$^{-3}$ for 
G55.3$-$1.5, when taking 5$\times$rms-noise as an upper limit and assuming the 
depth of the Faraday Screen is the mean of the projected major and minor 
axes. The resulting lower limits for the regular magnetic field strength 
along the line-of-sight is then about 8~$\mu$G for G40.9$-$4.2, and 
9~$\mu$G for G55.3$-$1.5. 

In the above discussions we have assumed uniform emissivity along the 
line-of-sight. For a local emissivity enhancement (see \citealt{srwe08} for 
details) the distance to the Faraday Screens is shorter, its size becomes 
smaller and its regular magnetic field increases. 

\subsection{Canals}

We see copious filamentary polarization minima in the $\lambda$6~cm survey maps,
 which were named ``canals" in previous low-frequency observations 
\citep[e.g.][]{hkd04}. The polarized intensity along the canals drops to 
typically 10\%--30\% compared to their surroundings. One of the most pronounced 
canals (see Fig.~\ref{sp12}) runs from $l=43\fdg6$, $b=-4\fdg8$ to $l=41\fdg0$, 
$b=-1\fdg0$. Another striking one is located at $l=33\degr$, $b=-3\fdg5$ 
(Fig.~\ref{sp12}). Many more canals appear towards the lower longitude regions, 
in particular the region between longitude of $10\degr$ and $13\degr$ 
(Fig.~\ref{sp34}). The length of the canals varies from tens of arcmin to 
several degrees. The width of the canals is about $5\arcmin$ to about 
$15\arcmin$, similar to the beam size.    

\begin{figure}[!htbp]
\centering
\resizebox{0.46\textwidth}{!}{\includegraphics[angle=-90]{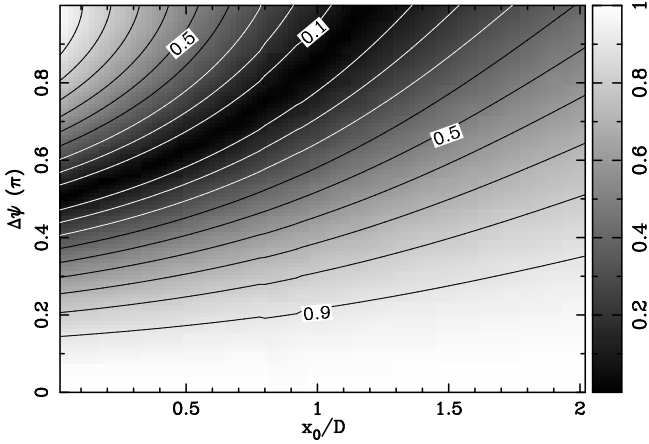}}
\caption{Depolarization factors versus polarization angle change ($\Delta\psi$) 
across the scale $x_0$ and the ratio of $x_0$ to the beam size $D$.}
\label{pex}
\end{figure}

The depolarization along the canals may be attributed to variations of 
polarization angles across the beam size $D$. Let the angles vary linearly by 
$\Delta\psi$ over a finite region $x_0$, i.e., 
$\psi(x)=\psi_0+\Delta\psi\frac{x}{x_0}$, where $|x|\leq x_0/2$. The 
depolarization factor $p$ can then be calculated following \citet{sbs+98} as, 
\begin{equation}
p=\frac{
\int\limits_0^{x_0/2D}\exp(ax^2)\cos(2\Delta\psi_Dx){\rm d}x+
\cos\Delta\psi\int\limits_{x_0/2D}^\infty\exp(ax^2){\rm d}x}
{\int\limits_0^\infty\exp(ax^2){\rm d}x},
\end{equation}
where $\Delta\psi_D=\Delta\psi D/x_0$ is the angle increment over 
the beam $D$, and $a=-4\ln2$. The results are shown in Fig.~\ref{pex}. 
The polarization angles around the canals in our maps show 
sharp variations, where $x_0\approx3\arcmin$ meaning $x_0/D\approx0.3$, and 
$0.5\pi\lesssim\Delta\psi\lesssim0.7\pi$. These parameters correspond  
depolarization factors of 0--0.3 in Fig.~\ref{pex}, which are consistent 
with the observations. 

\begin{figure*}
\centering
\resizebox{0.92\textwidth}{!}{\includegraphics[angle=-90]{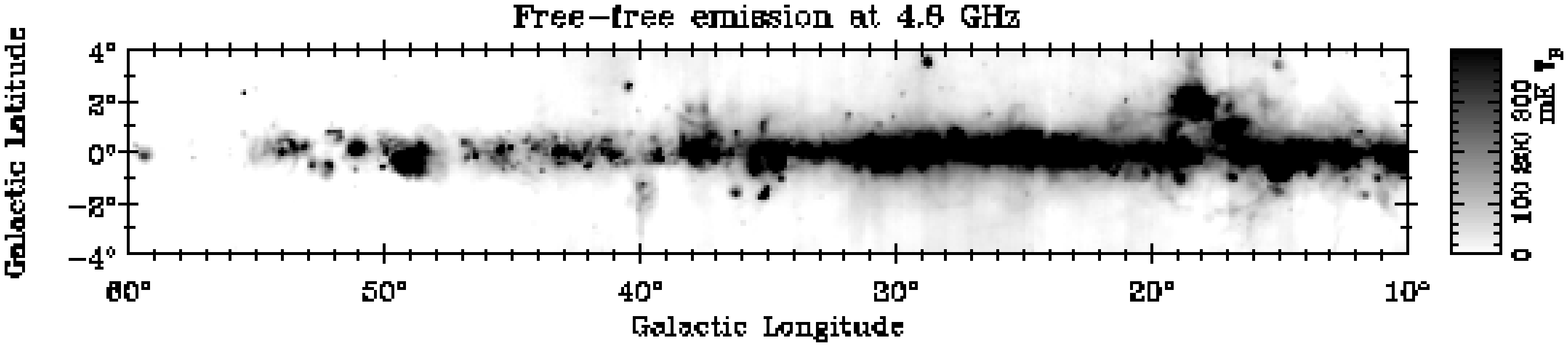}}\\[2mm]
\resizebox{0.92\textwidth}{!}{\includegraphics[angle=-90]{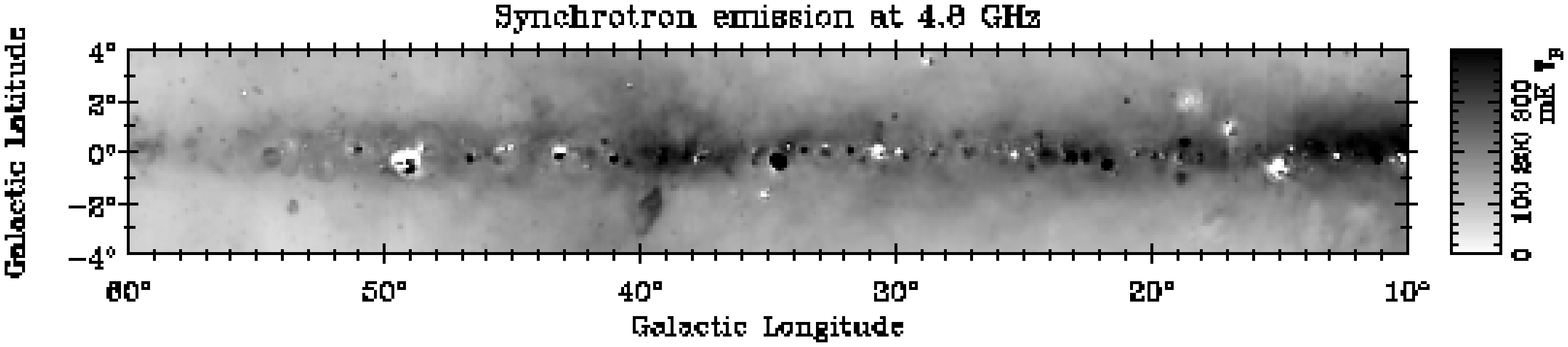}}
\caption{Decomposed thermal ({\it upper panel}) and non-thermal component 
({\it lower panel}) for the $\lambda$6~cm survey between 
$10\degr\leq l\leq60\degr$ based on the non-thermal spectral index model 
in Sect.~\ref{spmodel} and $\beta_{\rm ff}=-2.1$ for optically thin thermal emission.}
\label{thnth}
\end{figure*}

Variations of polarization angles can be either intrinsic or extrinsic 
\citep{hkd04}. At $\lambda$6~cm an angle change of $90\degr$ requires a Faraday 
depth of about 400~rad~m$^{-2}$, which is not supported by pulsar RMs for this 
area. Therefore an intrinsic scenario for the origin of the canals is likely, 
which means that they are caused by rapid variations of the polarization angle 
across the beam size. Thus canals delineate boundaries of polarized patches 
produced by turbulent magnetic field cells whose polarization angles differ. 
Whenever the angle difference matches the conditions for small depolarization 
factors as displayed in Fig.~\ref{pex} canals are expected to show up in the 
polarized intensity maps.

\section{Total intensity emission from the Galactic disk: thermal and 
non-thermal separation}

By combining the present $\lambda$6~cm survey section with the corresponding 
Effelsberg $\lambda$11~cm and $\lambda$21~cm maps we performed a decomposition 
of thermal and non-thermal emission components according to their different 
spectra. All maps were smoothed to $10\arcmin$ for this purpose. 

Observed total intensities T$(\nu)$ consist of several components: 
\begin{equation}
T(\nu)=T_{\rm gal}(\nu)+T_{\rm cmb}+T_{\rm egs}(\nu)+T_{\rm zero},
\end{equation}
where $T_{\rm cmb}=2.728\pm0.004$~K \citep{fcg+96} is the isotropic brightness 
temperature of the cosmic microwave background radiation. This component was 
already removed from the $\lambda$6~cm data by zero-level setting. The 
contribution from unresolved extragalactic sources $T_{\rm egs}(\nu)$ is only 
about several mK at $\lambda$6~cm \citep[e.g. ][]{rr88a} and thus can be 
ignored in view of the much higher Galactic emission in the Galactic plane. 
As stated before the $\lambda$6~cm maps miss Galactic large-scale emission and 
$T_{\rm zero}$ must be corrected by using other data. 

The emission from the Galaxy $T_{\rm gal}(\nu)$ can be written as 
\begin{equation}
T_{\rm gal}(\nu)=T_{\rm ff}(\nu)+T_{\rm syn}(\nu),
\end{equation} 
where $T_{\rm ff}(\nu)\propto\nu^{\beta_{\rm ff}}$ stands for the brightness 
temperature of free-free emission and 
$T_{\rm syn}(\nu)\propto\nu^{\beta_{\rm syn}}$ for synchrotron emission. 
It is known that for latitudes in the range $4\fdg5\leq|b|\leq5\degr$ the 
synchrotron emission clearly dominates the observed total intensity. We tie 
the zero-level of the $\lambda$6~cm data to the Effelsberg $\lambda$11~cm 
survey, which was corrected for missing large-scale emission (see 
\citealt{rfrr90} for details). The $\lambda$11~cm intensities for the area 
$4\fdg5\leq|b|\leq5\degr$ were extrapolated towards $\lambda$6~cm using the  
spectral index model devised in Sect.~\ref{spmodel}. The difference between 
the extrapolated and the original $\lambda$6~cm data were fitted linearly 
along latitudes and the offsets were added to the observed $\lambda$6~cm data. 
The assumption of a very small thermal emission fraction for the area 
$4\fdg5\leq|b|\leq5\degr$ was checked using the WMAP free-free emission 
template \citep{gbh+09}. By extrapolating the thermal component to 
$\lambda$11~cm we find that it amounts to 2\%--7\% of the total intensity. By 
neglecting this thermal emission the $\lambda$6~cm total intensity was 
underestimated up to 4~mK~$T_{\rm B}$, which is less than the $5\times\sigma$ 
level for total intensity.

For the separation of thermal and non-thermal emission we follow the method 
already used by \citet{pddg05} and others. The thermal fraction at frequency 
$\nu_0$ is calculated as 
\begin{equation}\label{fth}
\displaystyle{
f_{{\rm ff},\,\nu_0}=
\frac{1-\left(\frac{\nu}{\nu_0}\right)^{\beta_{\rm gal}-\beta_{\rm syn}}}
{1-\left(\frac{\nu}{\nu_0}\right)^{\beta_{\rm ff}-\beta_{\rm syn}}}}.
\end{equation} 
It is assumed that the combination of free-free emission and synchrotron 
emission results in a power-law, namely 
$T_{\rm gal}(\nu)\propto\nu^{\beta_{\rm gal}}$. The $\lambda$6~cm data together 
with the Effelsberg $\lambda$11~cm and $\lambda$21~cm data were used to derive 
the spectral index $\beta_{\rm gal}$. In fact good linear relations on  
logarithmic scales hold for almost all the pixels. The spectral index of 
free-free emission is fixed as $\beta_{\rm ff}=-2.1$, because the optical thin 
condition is in general satisfied in this wavelength range \citep{pddg05}. 
The spectral index of the synchrotron emission is modelled in 
Sect.~\ref{spmodel} according to the polarized intensities observed at K- and 
Ka-bands by WMAP. The spectral index towards the inner Galaxy is consistent 
with that obtained by \citet{rr88a} for the frequency range between 408~MHz and 
1420~MHz, which justifies the use of spectral indices derived in 
Sect.~\ref{spmodel} to the frequency range of 1.4~GHz to 4.8~GHz as well. For 
reliable results the three spectral indices have to satisfy 
$\beta_{\rm syn}<\beta_{\rm gal}<\beta_{\rm ff}$. If $\beta_{\rm gal}$ is 
smaller than the spectral index of synchrotron emission, we consider the entire 
emission as non-thermal. This situation happens at some spots out of the 
Galactic plane, where uncertainties in the zero-level setting have the largest 
influence on the spectral index. If $\beta_{\rm gal}>-2.1$, which happens for a 
few strong sources in the Galactic plane, the emission is completely attributed 
to the free-free radiation.

The so-separated thermal and non-thermal components for the inner Galactic 
plane are shown in Fig.~\ref{thnth}. The Effelsberg $\lambda$21~cm survey has 
latitude limits of $b=\pm4\degr$, so that we make the decomposition up to that 
latitude. It is obvious that the thermal emission component is more confined to 
the Galactic plane compared to the non-thermal emission. Note that some SNRs 
such as W~50 and SNR~G53.2$-$2.6 show up as distinct sources in the thermal 
emission, because their spectral indices are larger than that of the diffuse 
synchrotron emission. However, these individual objects do not influence our 
results, which focus on the large-scale diffuse emission. 

\begin{figure}[!htbp]
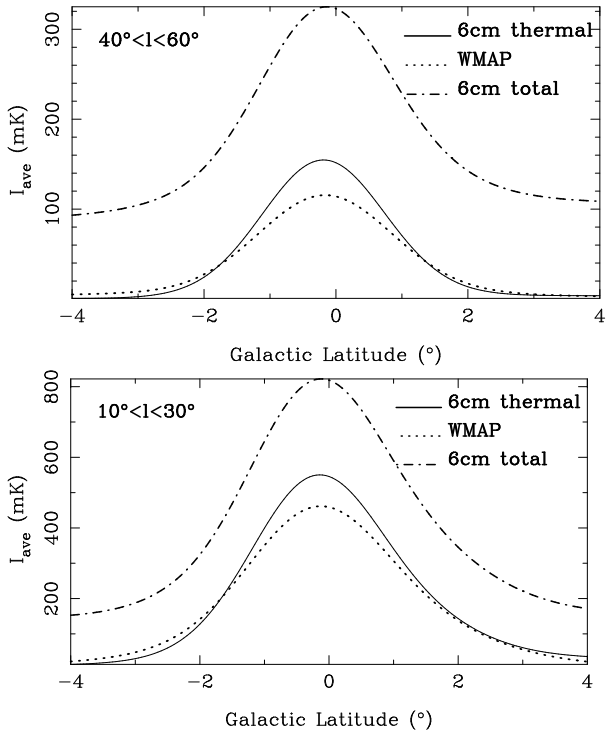

\includegraphics[width=4.8cm,angle=-90]{ith.fg.3.2deg.r1.ps}
\includegraphics[width=4.8cm,angle=-90]{ith.fg.3.2deg.r2.ps}
\caption{Galactic latitude profiles for the averaged thermal emission component 
at $\lambda$6~cm (solid lines), the averaged WMAP free-free emission (dotted 
line), and the baseline corrected $\lambda$6~cm total intensity (dot-dashed 
line).}
\label{ith}
\end{figure}

On average the thermal fraction is about 60\% at the Galactic plane for 
the longitude range of $10\degr\leq l\leq60\degr$. The thermal fraction 
decreases towards higher latitudes and drops to about 10\% for latitudes of 
$\pm2\degr$, and becomes about zero at latitudes beyond $\pm4\degr$. 
The thermal fraction reaches maximum values of about 78\% within the longitude 
range $20\degr<l<30\degr$ and latitude range of $|b|\leq 1\fdg5$. This fraction 
is similar as that 82\% reported by \citet{pddg05}. The corresponding maximum 
thermal fraction at 1.4~GHz is about 51\%. \citet{rr88b} estimated a fraction 
of about 40\% for the longitude range of $15\degr \leq l\leq50\degr$ at that 
frequency, whereas \citet{pddg05} obtained a value of 68\%. We ascribe this 
inconsistency to the different spectral indices used for the synchrotron 
component, which was $\beta_{\rm syn} = -2.7$ by \citet{pddg05}, but 
$\beta_{\rm syn} =-3.1$ for this region in the present study. The spectral 
index of synchrotron emission we derived from the WMAP K- and Ka-band polarized 
intensities in Sect.~\ref{spmodel} is quite probably the more reliable 
substitute for $\beta_{\rm syn}$. The maximum thermal fraction is about 60\% 
in the range of $36\degr<l<44\degr$ and $|b|\leq4\degr$, corresponding to about 
36\% at 1.4~GHz. This is consistent with the result obtained by \citet{add+10} 
based on a recombination line study when taking the uncertainties of this 
approach into account as discussed by \citet{add+10}.

We compared the thermal emission component derived with the WMAP five-year MEM 
free-free emission template \citep{gbh+09}. Both maps were smoothed to $2\degr$ 
angular resolution to eliminate the influence of small-scale structures. The 
profiles averaged for the longitude range of $10\degr<l<30\degr$ and of 
$40\degr<l<60\degr$ are shown in Fig.~\ref{ith}. Near the plane within 
$b=\pm2\degr$ the thermal emission we obtained is larger than that from the 
WMAP template by about 20\%. The inconsistency may stem for a small part from 
the uncertainties on recovering the large-scale emission at $\lambda$6~cm. 
There are, however, large uncertainties in the WMAP template \citep{gbh+09} 
towards the Galactic plane. In particular the influence of anomalous dust 
emission is unclear and needs further investigation. 

\begin{figure}[!htbp]
\includegraphics[width=4.8cm,angle=-90]{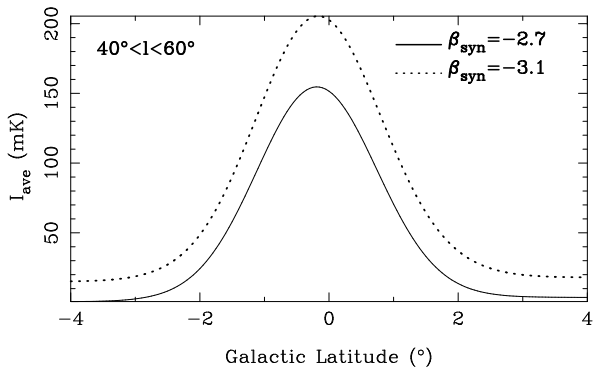}
\caption{Galactic latitude profiles for the averaged thermal emission component 
at $\lambda$6~cm derived with the spectral index $\beta_{\rm syn}=-2.7$ (solid 
lines) and $\beta_{\rm syn}=-3.1$ (dotted line).}
\label{th_err}
\end{figure}

Since the spectral index has particularly large uncertainties for the area 
$40\degr<l<60\degr$ we investigated its potential influence on the separation 
of thermal and non-thermal components. We used the most negative spectral index 
possible, $\beta_{\rm syn}=-3.1$, instead of $\beta_{\rm syn}=-2.7$, for 
calculating the missing large-scale emission to be added and repeated the 
separation with this spectral index. The difference is quite significant as 
displayed in Fig.~\ref{th_err}. However, using a spectral index of 
$\beta_{\rm syn}=-3.1$ results in significant thermal emission even at high 
latitudes. This clearly contradicts the results from WMAP. We conclude that the 
spectral index must be close to $\beta_{\rm syn}=-2.7$, which we used for the 
separation. 

\section{Summary}

We present $I$, $U$, $Q$ and $PI$ maps covering the section 
$10\degr\leq l\leq60\degr$ and $|b|\leq5\degr$ of the Sino-German 
$\lambda$6~cm survey of the Galactic plane. Observations and data processing 
were briefly described, and the restoration process to obtain correct 
zero-levels in $U$ and $Q$ by using the WMAP five-year K-band data, which needs
modification compared to the anti-centre survey sections because of different
polarization horizons at $\lambda$6~cm and K-band. Combining simulations by 
\citet{sr09} of the diffuse Galactic emission with RMs from pulsars we argue 
that the polarization horizon at $\lambda$6~cm is about 4~kpc in the Galactic 
plane, so that the observed polarized emission originates from the Sagittarius  
or the Scutum arm. At K-band the polarization horizon exceeds the dimension of 
the Galaxy. At absolute latitudes around $5\degr$, however, the differences in 
polarized emission at both wavelengths is largely reduced and almost the same 
volume is observed. 

There are various structures in the $\lambda$6~cm polarized intensity maps 
which have no counterpart in total intensity. We analyse the properties of 
polarized patches and conclude that they are caused by turbulent magnetic field 
cells. This demonstrates that polarization measurements have a great potential 
to retrieve the turbulent properties of Galactic magnetic fields. We show that 
the observed ``canals" occur when polarization angles vary by 
0.5$\pi$--0.7$\pi$ on scales of about $3\arcmin$ along the boundaries of 
polarized patches. Large depolarization ``voids' are regions where the 
polarized intensity drops close to zero. They result from Faraday Screens at 
about half the distance of the polarization horizon. The screens rotate 
background polarization so that it cancels foreground polarization of the same 
intensity. These Faraday Screens must host strong regular magnetic fields along 
the line-of-sight.

This $\lambda$6~cm survey section also enables high-frequency studies of known 
objects and the discovery of weak \ion{H}{II} regions and SNRs. We briefly 
discussed two newly identified \ion{H}{II} regions G12.8$-$3.6 and G56.7$-$0.6. 
We were able to trace the NPS extension closer towards the Galactic plane in 
polarization and total intensity than at longer wavelengths.

We separated thermal and non-thermal components for this Galactic plane region,
where an adjustment of the $\lambda$6~cm base-levels for total intensities was 
made by referring to the Effelsberg $\lambda$11~cm survey. For the separation 
process we decompose the total intensity according to the spectral index of 
thermal emission ($\beta_{\rm ff}=-2.1$) and of synchrotron emission. The 
synchrotron spectral index was deduced from WMAP K- and Ka-band polarization 
maps. At $\lambda$6~cm the  fraction of thermal emission 
in the plane is about 60\% on average.

We have demonstrated that the total intensity and polarized structures revealed 
in the $\lambda$6~cm survey maps provide new and important insights into the 
magneto-ionic properties of the Galaxy.
 
\begin{acknowledgements}
We like to thank the staff of the Urumqi Observatory for qualified 
assistance during the installation of the receiving system and the survey 
observations. In particular we are grateful to Otmar Lochner for the 
construction of the $\lambda$6~cm receiver, installation and commissioning. 
Maozheng Chen and Jun Ma helped with the installation of the $\lambda$6~cm 
receiving system and maintained it since 2004. We thank Prof. Ernst F\"urst for 
his support of the survey project and critical reading of the manuscript. 
The MPG and the NAOC supported the construction of the Urumqi $\lambda$6~cm 
receiving system by special funds. The Chinese survey team is supported by 
the National Natural Science foundation of 
China (10773016, 10833003, 10821061) and the National Key Basic Research 
Science Foundation of China (2007CB815403). XHS acknowledges financial support 
by the MPG and by Prof. Michael Kramer during his stay at MPIfR Bonn.
Finally we like to thank the referee for constructive comments, which helped
in particular to clarify the calibration procedures used.
\end{acknowledgements}

\bibliographystyle{aa}
\bibliography{journals}
\end{document}